\documentclass[12pt]{article}

\usepackage[allcolors=blue,colorlinks=true]{hyperref}

\usepackage{microtype}
\usepackage{graphicx}
\usepackage{subfigure}
\usepackage{booktabs} 
\usepackage{url}

\usepackage{mathtools}
\usepackage{microtype}
\usepackage{graphicx}
\usepackage{subfigure}
\usepackage{booktabs} 
\usepackage{amsmath,amssymb}
\usepackage{multirow}
\usepackage{multicol}
\usepackage{blindtext}
\usepackage{booktabs}
\usepackage{multirow}
\usepackage{siunitx}
\usepackage{caption}
\usepackage{doi}

\usepackage{paralist}
\usepackage{graphicx}
\usepackage{amsmath,amssymb}
\usepackage{color}
\usepackage{algorithm}
\usepackage{algorithmic}

\usepackage[dvipsnames]{xcolor}

\DeclareMathOperator{\bz}{\textbf z}

\usepackage{thmtools}

\DeclareMathOperator*{\argmin}{arg\,min}
\DeclareMathOperator*{\minimize}{minimize}
\DeclareMathOperator*{\subto}{subject\,to}
\newcommand{\vect}[1]{\boldsymbol{\mathbf{#1}}}
\DeclareMathOperator{\bTheta}{\mathbf \Theta}

\def\X{\mathbf X}
\def\U{\mathbf U}
\def\y{\mathbf y}

\def\V{\mathbf V}

\def\D{\mathbf D}

\def \curl {\boldsymbol\ell}
\newcommand{\bw}{\boldsymbol w}
\newcommand{\bbeta}{\boldsymbol \beta}
\def\bz{\mathbf z}
\def\Z{\mathbf Z}

\def\M{\mathbf M}
\def\N{\mathbf N}

\def\Q{\mathbf Q}

\usepackage[margin=1in]{geometry}

\usepackage[square,numbers]{natbib}
\usepackage[resetlabels]{multibib}
\usepackage{enumerate}
\usepackage{amsmath,latexsym,amssymb,graphicx,wrapfig,  enumitem, mathtools}
\allowdisplaybreaks[1]

\usepackage{footnote}
\usepackage{thm-restate}

\usepackage[utf8]{inputenc}
\usepackage[english]{babel}

\usepackage[dvipsnames]{xcolor}

\usepackage{etex}
\usepackage[etex=true,export]{adjustbox}

\usepackage{natbib}
\setcitestyle{round}
\usepackage{multirow}
\usepackage{multicol}
\usepackage{siunitx}

\makesavenoteenv{tabular}
\makesavenoteenv{table}

\makeatletter
\renewcommand*{\@fnsymbol}[1]{\@alph{#1}}
\makeatother

\bibpunct{(}{)}{;}{a}{}{,}

\usepackage{setspace}

\begin{document}


\title{\bf Supervised Convex Clustering}
\author{ Minjie Wang\thanks{Department of Statistics, Rice University, Houston, TX},\hspace{.2cm}
Tianyi Yao\footnotemark[1],\hspace{.2cm}
and Genevera I. Allen\thanks{Departments of Electrical and Computer Engineering, Statistics, and Computer Science, Rice University, Houston, TX} \textsuperscript{,}\thanks{Jan and Dan Duncan Neurological Research Institute, Baylor College of Medicine, Houston, TX}}
\date{}
\maketitle

\begin{abstract}
 Clustering has long been a popular unsupervised learning approach to identify groups of similar objects and discover patterns from unlabeled data in many applications. Yet, coming up with meaningful interpretations of the estimated clusters has often been challenging precisely due to its unsupervised nature. Meanwhile, in many real-world scenarios, there are some noisy supervising auxiliary variables, for instance, subjective diagnostic opinions, that are related to the observed heterogeneity of the unlabeled data. By leveraging information from both supervising auxiliary variables and unlabeled data, we seek to uncover more scientifically interpretable group structures that may be hidden by completely unsupervised analyses. In this work, we propose and develop a new statistical pattern discovery method named Supervised Convex Clustering (SCC) that borrows strength from both information sources and guides towards finding more interpretable patterns via a joint convex fusion penalty. We develop several extensions of SCC to integrate different types of supervising auxiliary variables, to adjust for additional covariates, and to find biclusters. We demonstrate the practical advantages of SCC through simulations and a case study on Alzheimer’s Disease genomics.
Specifically, we discover new candidate genes as well as new subtypes of Alzheimer’s Disease that can potentially lead to better understanding of the underlying genetic mechanisms responsible for the observed heterogeneity of cognitive decline in older adults. 
\end{abstract}

\noindent%

\newpage


\doublespacing

\section{Introduction}
\label{sec:intro}

Clustering is an unsupervised learning approach that seeks to find groups of objects which are similar to each other. Despite successes in applying clustering in many fields such as genomics, online advertising, and text mining, coming up with meaningful interpretations of the estimated clusters has often been challenging precisely due to its unsupervised nature. Currently in practice, most people cluster data in a fully unsupervised manner and then interpret the clustering results via some outcomes of interest or other meta-data that help to validate the clusters. We call these ``supervising auxiliary variables".
Our goal is to use the supervising auxiliary variables as part of the clustering procedure itself to help guide towards finding more accurate and interpretable clusters.

Let us consider our motivating case study on the clinical genomics of Alzheimer's Disease (AD), which will be discussed in more detail in Section~\ref{case_study}. All individuals in this case study experience cognitive decline as they age, yet cognitive abilities of some subjects decline at a much faster rate than others and there is a large degree of heterogeneity in cognitive skills. Understanding such heterogeneity in cognitive decline can better elucidate the underlying genetic mechanisms responsible for AD and other dementias. However, apart from a handful of well-known genes such as APOE, little is known about the genomics of AD and of cognitive decline in older adults. It is common for people to study this by clustering the subjects and validating the results using additional information such as the clinical diagnosis or cognitive test scores. However, compelling evidence of genetic subtypes for AD and cognitive decline has not been found. We propose to use the additional meta information such as the cognitive test scores or clinical diagnosis directly to help us find better and more interpretable clusters, and hence, shed new light on the genetic basis responsible for onset of AD and dementia.  In many other genomics studies, additional clinical information or survival times are available and can thus be used as supervising auxiliary variables to guide clustering.

Yet, making use of the supervising auxiliary variables to help guide towards finding groups presents several major challenges. First, due to human subjectivity and measurement errors, these clinical outcomes are noisy and thus cannot be fully trusted as ground-truth outcomes or labels. Specifically, we do not know how much supervision we should get from these noisy supervising auxiliary variables when clustering unlabeled data. Second, these clinical outcomes can be of different data types. For example, diagnostic opinions assigned by biologists can be categorical while survival time is censored data.

To address these challenges, in this paper, we seek to leverage information from both supervising auxiliary variables, usually of different types, and unlabeled data to uncover more scientifically interpretable group structures that may be hidden in completely unsupervised analyses of data.  Our approach is distinct from supervised learning, which treats these outcomes as ground truth to make scientific discoveries, as the supervised approaches fail to exploit the unlabeled data to uncover group structures. In addition, these supervising auxiliary variables are different from the outcomes or labels in supervised learning in that they are largely noisy and thus cannot be fully trusted. Our method is also distinct from unsupervised approaches as we make better use of these potentially meaningful supervising auxiliary variables to understand the true underlying group structures. Although these supervising auxiliary variables are not true labels or outcomes themselves, they have loose relationship with the group structure in the data and hence indicate some forms of observed heterogeneity of the unlabeled data.

Though supervised clustering has not been widely studied, there is a plethora of literature on semi-supervised clustering specific to the nature of the outcome variables.  For partially labeled data, \cite{basu2002semi} proposed to modify the objective function of $k$-means to compute initial cluster centroids by incorporating such labels. 
In this case, labeled observations are always assigned to their known cluster. However, such approaches usually assume the labels are perfect and require prior knowledge of total number of clusters. 
In other scenarios, people incorporate prior information on pairwise (must-link or cannot-link) constraints that dictate whether two data points must be clustered in the same group or not.
To take those constraints into account, 
\cite{basu2004active,xing2003distance,bar2003learning} modified the objective function of existing clustering methods or the distance metric in the distance-based clustering method. Still, those methods require that our presumed knowledge of the constraints is correct. 

On the other hand, many have tried to improve the interpretability of clustering results by only using features related to some supervised outcome. Specifically, \citet{bair2004semi,koestler2010semi,gaynor2013identification} discarded or down-weighted unimportant features by univariate filtering associated with the outcome and then performed clustering on the meaningful features. However, such methods might neglect features that are weakly associated with the outcome variable but differ across clusters. Above all, these semi-supervised clustering approaches 
make full use of the noisy outcome variable without adjustment. Yet, our goal is to leverage information from both supervising auxiliary variables, which may be imperfect, and unlabeled data to obtain more interpretable group structures.

Another line of work focuses on semi-supervised classification which uses both labeled and unlabeled data to improve the performance of classifiers. The semi-supervised SVM minimizes the objective function by examining all possible label combinations of unlabeled data points and then finds low density regions that the decision boundary could pass through \citep{chapelle2007branch,chapelle2006continuation,yuille2003concave}.
The cluster-then-label techniques first find clusters of high density regions in data space by clustering. A standard supervised learner is then applied to find a separating decision boundary that passes through the low density regions \citep{chapelle2005semi,gan2013using}.
A somewhat related line of work proposes some classification methods that can handle noisy or missing labels \citep{bi2014multilabel}. 
\citet{angluin1988learning} proposed random classification noise model which assumes each label  is  flipped independently with some probability less than 0.5. Other popular approaches include using losses that are robust to the presence of noisy labels such as 0-1 loss \citep{manwani2013noise} or modifying surrogate loss functions that approximate the 0-1 loss via a convex function \citep{natarajan2013learning}. We refer the reader to the survey by \citet{frenay2013classification}.
Our method is fundamentally different from both semi-supervised classification or classification with noisy labels methods in that the major task of these methods is supervised learning (prediction) while our goal is to find groups using the supervising auxiliary variables.

We seek to develop a unified, convex formulation of supervised clustering based on increasingly popular convex clustering methods. \citet{pelckmans2005convex, lindsten2011just, hocking2011clusterpath} studied a convex formulation of clustering that achieves agglomeration through a convex fusion penalty. Due to this convex formulation, it enjoys nice statistical properties such as global optimal solutions, stable solutions to small perturbation of data \citep{pelckmans2005convex,chi2017convex} and statistical consistency \citep{radchenko2017convex,tan2015statistical}. Recently, to address the expensive computation of convex clustering, \citet{chi2015splitting,weylandt2019dynamic} developed fast and efficient algorithms to solve the convex clustering problem and yield full regularization paths. Further, convex clustering has been extended to many applications such as convex biclustering \citep{chi2017convex}, which allows for clustering features simultaneously, and closely related to our work, recently \citet{wang2019integrative} adopted the convex clustering approach to perform integrative clustering for high dimensional mixed, multi-view data.


In this paper, we propose and develop a new statistical pattern discovery method named Supervised Convex Clustering (SCC) that borrows strength from both the unlabelled data and supervising auxillary variables to find more interpretable patterns. 
Specifically, we develop an optimization problem defining our method that consists of three parts: an unsupervised loss for the unlabeled data, a supervised loss that incorporates the supervising auxiliary variable, and a joint convex fusion penalty that forces the group structure of the unlabeled data and the supervising auxiliary variable to be the same. Our method, to the best of our knowledge, is the first to perform supervised clustering that directly uses supervising auxiliary variables to help cluster unlabeled data. 

\section{Supervised Convex Clustering Method}
In this section, we propose and develop our supervised convex clustering method for different types of supervising auxiliary variables. Then we discuss some practical considerations for applying our method and develop an adaptive approach to adjust for additional covariates.

\subsection{General Model \& Formulation}
Let $(y_i,\mathbf{X}_{i\cdot})$ denote the pair of supervising auxiliary variable $y_i$ and feature vector $\mathbf{X}_{i\cdot}\in\mathbb{R}^{p}$ for the $i^{\text{th}}$ observation, $i\in\{1,\hdots,n\}$. Let $(\theta_i,\mathbf{U}_{i\cdot})$ denote the corresponding pair of supervising auxiliary variable centroid $\theta_i$ and data centroid $\mathbf{U}_{i\cdot}\in\mathbb{R}^{p}$ for the $i^{\text{th}}$ observation. Let $\mathbf{Z}_{i\cdot} \in \mathbb{R}^{d}$ denote the additional covariates associated with the supervising auxiliary variable for the $i^{\text{th}}$ observation and $\bbeta \in \mathbb{R}^{d}$ is the corresponding vector of coefficients. Define $g(\cdot)$ to be the appropriate link function for a Generalized Linear Model (GLM) whose exact form depends on the data type of the supervising auxiliary variable $\mathbf{y}$ (continuous, skewed-continuous, binary, and count-valued, among others). For example, if $\y$ is count-valued data, $g(\cdot)$ can be  the log-link.
Define $C: i \to k$ to be a function which maps from the observation indices $i \in \{1,\hdots,n\}$ to cluster labels $k \in \{1,\hdots,K\}$. Then for $i \in \{1,\hdots,n\}$, we consider the following data-generating model:
\begin{align*}
    \mathbf{X}_{i\cdot} &= \mathbf{U}_{i\cdot} + \mathbf{E}_{i\cdot}, \:\:\:\:\mathbf{E}_{i\cdot}\sim \text{MVN}(\mathbf{0}_p, \sigma^2\mathbf{I}) \\
    g(\mathbb{E}[y_i|\mathbf{Z}_{i\cdot}]) &= \theta_{i} + 
    \mathbf{Z}_{i\cdot}^T\vect{\beta} \\
    \mathbf{U}_{i\cdot} &= \mathbf{U}_{j\cdot}, \hspace{2mm} \theta_i = \theta_j, \hspace{10mm} \text{if} \hspace{2mm} C(i) = C(j) = k.
\end{align*}
This model assumes the unlabeled data follows a group mean (centroid) plus noise model.  The supervising auxillary variable, adjusted for covariates $\Z_{i\cdot}$, follows a GLM whose mean has the same group structure as the unlabeled data.  


We propose to fit this model by formulating a convex optimization problem based on convex clustering.  Let $\ell(\cdot)$ be the negative log-likelihood or loss function for the particular generalized linear model associated with $g(\cdot)$.  Our Supervised Convex Clustering method is hence the solution to the following optimization problem:
\begin{align}\label{eq:1}
\begin{split}
    \minimize_{\mathbf{U}\in\mathbb{R}^{n\times p}, \vect{\theta} \in \mathbb{R}^{n}, \vect{\beta}  \in \mathbb{R}^{d}} &\pi_{\mathbf{X}} \cdot \frac{1}{2} \sum_{i=1}^n ||\mathbf{X}_{i.}-\mathbf{U}_{i.}||_2^2 + \pi_{\mathbf{y}} \cdot \sum_{i=1}^n  \ell(y_i;\theta_i+\mathbf{Z}_{i.}^T \vect{\beta})  \\
    & + \lambda \sum_{1 \leq i < j \leq n} w_{ij}\Big|\Big|\left[
\begin{array}{c}
\theta_i  \\
\mathbf{U}_{i\cdot}
\end{array}
\right] - \left[
\begin{array}{c}
\theta_j  \\
\mathbf{U}_{j\cdot}
\end{array}
\right]\Big|\Big|_2.  
\end{split} 
\end{align}
Here, $\pi_\X$ and $\pi_{\mathbf{y}}$ are fixed inputs by the user in advance; $\lambda$ is a non-negative tuning parameter; and, $w_{ij}$ are non-negative user-specific fixed inputs.

Our optimization problem can be thought of as an extension of convex clustering that incorporates supervised data. One way to interpret this is that we have a loss function for the unlabeled data, a loss function for the supervising auxiliary variable, and a new joint convex fusion penalty that connects the supervised and unsupervised parts.
Specifically, we employ a joint group-lasso fusion penalty on the concatenated centroid $\begin{bmatrix} \vect{\theta} &  \U \end{bmatrix}$ that forces the group structure (cluster assignment) of the $i^{\text{th}}$ row of $\U$ to be the same as that of $\vect{\theta}$. Similar to convex clustering, our joint group-lasso-type fusion penalty encourages the differences in the rows to be shrunk towards zero, inducing a clustering behavior. Here, $\lambda$ is a positive tuning parameter which regulates both the cluster assignment and number of clusters. When $\lambda$ equals zero, each observation forms its own cluster centroid. As $\lambda$ increases, the fusion penalty encourages the rows of concatenated centroid  $\begin{bmatrix} \vect{\theta} &  \U \end{bmatrix}$ to merge together, forming clusters. We say that subjects with the same centriods belong to the same cluster, which means, $\X_{i.}$ and  $\X_{j.}$ have the same cluster membership if $\U_{i.} = \U_{j.}$ and $\theta_i = \theta_j$. As $\lambda$ is sufficiently large, all the rows of concatenated centroid  $\begin{bmatrix} \vect{\theta} &  \U \end{bmatrix}$ coalesce to a single cluster centroid.  Our joint fusion penalty is novel as it puts together the centroids for both the unlabeled data and the supervising auxiliary variable, forcing them to have the same group structure. In this way, our method borrows strength from both information sources and yields the same cluster assignment for similar observations. 
The weight $w_{ij}$, which manifests pairwise affinity, is a user-specific input which will be discussed in detail in Section \ref{practical}.

The two loss functions, for the unsupervised and supervised part respectively,  are weighted by $\pi_{\mathbf{X}}$ and $\pi_{\mathbf{y}}$ respectively, which are deterministic parameters for fixed data. 
The user can directly specify these weights according to how much they want to weight the unsupervised 
and supervised part. But if one wants these to be completely data-driven determined hyper-parameters, we have found that setting $\pi_{\mathbf{X}}$ and $\pi_{\mathbf{y}}$ to be inversely proportional to the null deviance evaluated at the loss-specific center, i.e.,  $\pi_{\mathbf{X}} = \frac{1}{\frac{1}{2}||\mathbf{X}-\bar{\mathbf{X}}||_F^2}$, $\pi_{\mathbf{y}}  = \frac{1}{\ell(\mathbf y, \mathbf{\tilde y})}$, performs well in practice. 
Here $\mathbf{\tilde y}$ denotes the loss-specific center for loss $\ell(\cdot)$ as discussed in \cite{wang2019integrative}. We use such $\pi$'s so that two losses are evaluated at the same scale in the objective function. 
Suppose we remove the second term in the objective function ($\pi_{\mathbf y} = 0$), we get a fully unsupervised method. Similarly, if we remove the first term ($\pi_{\X} = 0$), we perform convex clustering on the supervising auxiliary variable alone.



We employ different loss functions to account for the unlabeled data and supervising auxiliary variables of different data types. Notice that here we assume the data matrix $\X$ to follow Gaussian distribution and use Euclidean distances as the loss function, but one could easily generalize it to any convex losses for non-Gaussian data $\X$. The general loss $\ell(.)$ is a convex 
function whose specific form depends on the data type of $\mathbf{y}$.  For example, $\ell$ can be the negative log-likelihood for any common generalized linear models such as Gaussian, logistic, log-linear (Poisson), negative binomial. For supervising auxiliary variable which is categorical or survival data, the general form above does not apply and we need some minor changes to the formulation. We specify these in the next subsections as special cases.

\subsection{Special Case: Categorical Supervising Auxiliary Variable}
We model a categorical supervising auxiliary variable with $K$ classes using the multinomial loss. To facilitate this, we first transform the supervising auxiliary variable into dummy variables $\mathbf Y \in\mathbb{R}^{n \times K}$ where $\mathbf Y_{ik} = 1$ if subject $i$ belongs to the $k^{\text{th}}$ class. 

By construction, we employ negative log-likelihood of multinomial distribution as loss $\ell(\cdot)$ and similarly denote  $\vect{\bTheta} \in\mathbb{R}^{n \times K}$ as the centroid matrix for supervising auxiliary variable  $\mathbf Y$; this gives supervised convex clustering for categorical supervising auxiliary variables.
{
\centering
\scalebox{0.85}{\parbox{1\linewidth}{%
\begin{align*}
\begin{split}
    \minimize_{\mathbf{U}\in\mathbb{R}^{n\times p}, \bTheta \in\mathbb{R}^{n\times K}, \bbeta_k \in \mathbb{R}^{d} } &\pi_{\mathbf{X}} \cdot \frac{1}{2}  \sum_{i=1}^n ||\mathbf{X}_{i.}-\mathbf{U}_{i.}||_2^2  + \pi_{\mathbf{y}} \cdot \sum_{i=1}^n \bigg \{ \sum_{k=1}^K - y_{ik} (\theta_{ik}+\bz_i^T \bbeta_k) + \log (\sum_{k=1}^K e^{\theta_{ik}+\bz_i^T \bbeta_k} ) \bigg \}    \\
    &+ \lambda \sum_{1 \leq i < j \leq n} w_{ij}\Big|\Big|\left[
\begin{array}{c}
\bTheta_{i\cdot} \\
\mathbf{U}_{i\cdot}
\end{array}
\right] - \left[
\begin{array}{c}
\bTheta_{j\cdot}  \\
\mathbf{U}_{j\cdot}
\end{array}
\right]\Big|\Big|_2
\end{split}
\end{align*}
}}
\par
}

 We enforce a joint fusion penalty on the rows of concatenated centroid $\begin{bmatrix} \vect{\bTheta} &  \U \end{bmatrix}$ to yield shared group structure between two sources. Note as in the regular multinomial regression problem, this parameterization of $\bbeta_k$ is not identifiable as the value of objective function would not change if we change $\bbeta_k$ with $\bbeta_k + c$ for all $k$. To address this issue, we add the constraint $\sum_{k=1}^K \bbeta_k = \mathbf 0$ as discussed in \citet{zhu2004classification}.

\subsection{Special Case: Censored Survival Time as Supervising Auxiliary Variable}

Following the Cox Proportional Hazards (CPH) model, suppose we observe  data with survival time $(\mathbf{x}_i, \delta_i,t_i)$ where $\delta_i$ denotes the censoring indicator and $t_i$ refers to the censoring time. For each $i=1,\hdots,n$, denote $R_{t_i}$ as the risk set of individuals who are alive and in the study at time $t_i$, $R_{t_i} = \{j: t_j \geq t_i \}$. All survival times are assumed to be unique; for tied times, we use Breslow's approximation. The supervised convex clustering problem can be formulated as follows:

{
\centering
\scalebox{0.85}{\parbox{1\linewidth}{%
\begin{align*}
\begin{split}
    \minimize_{\mathbf{U}\in\mathbb{R}^{n\times p}, \vect{\theta}\in \mathbb{R}^{n},\vect{\beta} \in \mathbb{R}^{d} } &\pi_{\mathbf{X}} \cdot \frac{1}{2}  \sum_{i=1}^n ||\mathbf{X}_{i.}-\mathbf{U}_{i.}||_2^2  + \pi_{\mathbf{y}} \cdot \bigg[  -\sum_{i=1}^n \delta_i (\theta_i + \bz_i^T \bbeta)  + \sum_{i=1}^n \delta_i \log \big \{ \sum_{j \in R(t_i)} \exp( \theta_j + \bz_j^T \bbeta)   \big \} \bigg]  \\
    & + \lambda \sum_{1 \leq i < j \leq n} w_{ij}\Big|\Big|\left[
\begin{array}{c}
\theta_i  \\
\mathbf{U}_{i\cdot}
\end{array}
\right] - \left[
\begin{array}{c}
\theta_j  \\
\mathbf{U}_{j\cdot}
\end{array}
\right]\Big|\Big|_2.
\end{split}
\end{align*}
}}
\par
}

Here, we interpret the supervising auxiliary variable centroid $\vect{\theta}$ as the hazard rate for the $i^{\text{th}}$ subject. A larger $\vect{\theta}$ indicates that the event is more likely to be observed for the subject. Hence, one interpretation of supervised convex clustering is that we are finding groups that have different hazard rates for survival.

\subsection{Supervised Convex Biclustering}\label{scc-biclust}


To allow for grouping observations and features simultaneously, we extend our method to supervised convex biclustering based on the approach discussed by \citet{chi2017convex}.  The supervised convex biclustering problem can be formulated as follows:
\begin{align*} 
\begin{split}
    \minimize_{\mathbf{U}\in\mathbb{R}^{n\times p}, \vect{\theta}\in \mathbb{R}^{n}, \vect{\beta} \in \mathbb{R}^{d}} &\pi_{\mathbf X} \cdot \frac{1}{2} \sum_{i=1}^n ||\mathbf{X}_{i.}-\mathbf{U}_{i.}||_2^2  +  \pi_{\mathbf y}  \cdot  \sum_{i=1}^n  \ell(y_i;\theta_i+\mathbf{Z}_{i.}^T \vect{\beta})\\ &+ \lambda \sum_{1 \leq i < i
     \leq n} w_{ii'}\Big|\Big|\left[
\begin{array}{c}
\theta_i  \\
\mathbf{U}_{i\cdot}
\end{array}
\right] - \left[
\begin{array}{c}
\theta_{i'}  \\
\mathbf{U}_{i'\cdot}
\end{array}
\right]\Big|\Big|_2  + \lambda \sum_{1 \leq j < j' \leq p} \tilde w_{jj'}\Big|\Big| \mathbf{U}_{\cdot j} - \mathbf{U}_{\cdot j'}
 \Big|\Big|_2.
\end{split}
\end{align*}

The row-wise fusion penalty fuses the rows of concatenated  centroid $\begin{bmatrix} \vect{\theta} &  \U \end{bmatrix}$ while the column-wise fusion penalty fuses the columns of $\U$.  Note that we fuse observations based on the data matrix and the supervising auxiliary variable while clustering only the features in the data matrix. For the choice of the weights $w_{ii'}$ and $w_{jj'}$, we refer the reader to \citet{chi2017convex} and will discuss this in detail in Section \ref{practical}.  The two penalties jointly achieve a checkerboard pattern that illustrates the associations between groups of subjects guided by the supervised auxillary variable and groups of features that distinguish the subjects.

\subsection{Doubly-supervised Convex Biclustering}

In some cases, we observe meta-data or supervising auxiliary information for both the rows (subjects) and the columns (features) of the unlabeled data. Suppose, besides supervising auxiliary variable for the $i^{\text{th}}$ subject, $y_i$, we observe another supervising auxiliary variable for the $j^{\text{th}}$ feature, denoted as $\tilde y_j$. Denote $\vect{\tilde \theta} \in \mathbb R^p$ as the cluster centroid for the supervising auxiliary variable $\mathbf{\tilde y}  \in \mathbb R^p$; denote $\mathbf{\tilde Z}\in\mathbb{R}^{p \times \tilde d}$ as the additional covariates that are associated with outcome $\mathbf{\tilde y}$, and denote $\tilde \bbeta \in \mathbb{R}^{\tilde d}$ as the corresponding vector of coefficients.
Let $\tilde \ell(\cdot)$ be the negative log-likelihood or loss function for the supervising auxiliary variable $\vect{\tilde y}$. Let $\pi_{\mathbf {\tilde y}}$ be fixed input. The doubly-supervised convex clustering problem can be formulated as follows:

{
\centering
\scalebox{0.9}{\parbox{1\linewidth}{%
\begin{align*} 
\begin{split}
    \minimize_{\substack{\mathbf{U}\in\mathbb{R}^{n\times p}, \vect{\theta} \in \mathbb{R}^{n}, \vect{\beta} \in \mathbb{R}^{d}, \\ \vect{\tilde \theta} \in \mathbb{R}^{p}, \tilde \bbeta \in \mathbb{R}^{\tilde d}}} &\pi_{\mathbf X} \cdot \frac{1}{2} \sum_{i=1}^n ||\mathbf{X}_{i.}-\mathbf{U}_{i.}||_2^2  +  \pi_{\mathbf y}  \cdot  \sum_{i=1}^n  \ell(y_i;\theta_i+\mathbf{Z}_{i.}^T \vect{\beta})   +  \pi_{\mathbf  {\tilde  \y}}  \cdot  \sum_{j=1}^p \tilde \ell(\tilde y_j;\tilde \theta_j+ \mathbf{\tilde Z}_{j.}^T \vect{\tilde \beta})\\ 
    &+ \lambda \sum_{1 \leq i < i
     \leq n} w_{ii'}\Big|\Big|\left[
\begin{array}{c}
\theta_i  \\
\mathbf{U}_{i\cdot}
\end{array}
\right] - \left[
\begin{array}{c}
\theta_{i'}  \\
\mathbf{U}_{i'\cdot}
\end{array}
\right]\Big|\Big|_2  + \lambda \sum_{1 \leq j < j' \leq p} \tilde w_{jj'}\Big|\Big|\left[
\begin{array}{c}
\tilde \theta_j  \\
\mathbf{U}_{\cdot j}
\end{array}
\right] - \left[
\begin{array}{c}
\tilde \theta_{j'}  \\
\mathbf{U}_{\cdot j'}
\end{array}
\right]\Big|\Big|_2.
\end{split}
\end{align*}
}}
\par
}
Our doubly-supervised convex clustering can be interpreted as performing supervised clustering on both the subjects and features of the unlabeled data that directly uses two sources of supervising auxiliary variables.

\subsection{Algorithm}
In this subsection, we propose an algorithm to solve our supervised convex clustering problem. Since there are more than two separate functions in our problem, the most common approach is to use multi-block ADMM \citep{lin2015global,deng2017parallel}, which decomposes the original problem into several smaller and easier sub-problems.

Denote $\D \in \mathbb R^{ |\mathcal E | \times n} $ as  the  directed  difference  matrix  corresponding  to  the  non-zero  fusion weights. We can recast the supervised convex clustering problem  \eqref{eq:1} as the equivalent constrained optimization problem. 
\begin{align*}
    &\minimize_{\mathbf{U}, \vect{\theta}, \vect{\beta},\V} \hspace{5mm} \pi_{\mathbf{X}} \cdot \frac{1}{2}||\mathbf{X}-\mathbf{U}||_F^2 + \pi_{\mathbf{y}} \cdot \ell(\mathbf{y};\vect{\theta}+\mathbf{Z}\vect{\beta}) + \lambda \underbrace{ \bigg(\sum_{l \in \mathcal E} w_l \|\V_{l.}\|_2\bigg)}_{P(\V;\bw)} \\    
&\subto \hspace{5mm} \D \begin{bmatrix} \vect{\theta} & \U  \end{bmatrix} - \V = \mathbf 0 
\end{align*}

To facilitate the constraints above in matrix-form, we concatenate the supervising auxiliary variable centroid vector $\vect{\theta}\in\mathbb{R}^{n}$ and the data centroid matrix $\mathbf{U}\in\mathbb{R}^{n\times p}$ into the aggregated centroid matrix $\begin{bmatrix} \vect{\theta} &  \U \end{bmatrix} \in\mathbb{R}^{n\times(p+1)}$. We then introduce an auxiliary variable $\mathbf{V} = 
\begin{bmatrix} \mathbf{V}_{\vect{\theta}} &  \mathbf{V}_{\mathbf{U}}  \end{bmatrix}
\in\mathbb{R}^{|\mathcal{E}|\times (p+1)}$ containing the pairwise differences between connected rows of the aggregated centroid matrix $\begin{bmatrix} \vect{\theta} &  \U \end{bmatrix}$.  The constraints can now be written as  $\D \begin{bmatrix}\vect{\theta}  & \U  \end{bmatrix} - \V = \mathbf 0$. Also, we replace the squared Euclidean distances with the squared Frobenius norm to facilitate matrix-version algorithm. We can yield the augmented Lagrangian and apply multi-block ADMM to solve our supervised convex clustering problem. 

Further, note that both the $\vect{\theta}$ and $\bbeta$ sub-problems generally do not have analytical closed-form solutions for arbitrary loss function $\ell$. Hence we need to apply an inner optimization routine with nested iterative updates to solve the sub-problem until full convergence, which is computationally intensive. To address this and speed up computation, we adopt the generalized multi-block ADMM with inexact sub-problem approach by \citet{wang2019integrative} and take a one-step descent update to solve the sub-problem approximately.  For differentiable loss $\ell$, we take a one-step gradient descent update by applying linearized multi-block ADMM to the $\vect{\theta}$ or $\bbeta$ sub-problem for each iteration. This gives Algorithm~\ref{alg:scc-alg}, a multi-block ADMM  algorithm to solve supervised  convex  clustering with differentiable loss. Similarly, for non-differentiable distance-based loss $\ell$, we can introduce a new block for the non-smooth function $\ell$ and apply multi-block ADMM with simple closed-form solutions for each primal variable update. The dual variable is denoted by $\Q$.

\begin{algorithm}[H]
	\caption{Multi-block ADMM algorithm for supervised convex clustering with differentiable loss $\ell$}
    \label{alg:scc-alg}
	\begin{algorithmic}

		\WHILE{not converged}

		\STATE 	$\mathbf{U}^{(k+1)}= (\pi_{\mathbf{X}} \cdot \mathbf{I}+\rho\mathbf{D}^T\mathbf{D})^{-1}\Big(\pi_{\mathbf{X}}\mathbf{X}+\rho\mathbf{D}^T(\mathbf{V}_{\mathbf{U}}^{(k)} - \mathbf{Q}_{\mathbf{U}}^{(k)})\Big)$ 

        \STATE $\vect{\theta}^{(k+1)} = \vect{\theta}^{(k)} - t_k \big(  \pi_{\mathbf{y}} \cdot \nabla \ell(\mathbf{y};\vect{\theta}^{(k)}+\mathbf{Z}\vect{\beta}^{(k)}) + \rho \mathbf{D}^T (\mathbf{D}\vect{\theta}^{(k)} - \mathbf{V}_{\vect{\theta}}^{(k)} + \mathbf{Q}_{\vect{\theta}}^{(k)} ) \big)$  
        
        \STATE $\vect{\beta}^{(k+1)}= \vect{\beta}^{(k)} - t_k  \nabla \ell(\mathbf{y}; \vect{\theta}^{(k+1)}+\mathbf{Z}\vect{\beta}^{(k)})$ \;
		
		\STATE $\mathbf{V}^{(k+1)} = \text{prox}_{\lambda/\rho P(\cdot; \bw )} (\mathbf{D}\begin{bmatrix} \vect{\theta}^{(k+1)} &  \U^{(k+1)} \end{bmatrix} + \Q^{(k)})$  
		
		\STATE $\mathbf{Q}^{(k+1)} = \mathbf{Q}^{(k)} + (\mathbf{D}\begin{bmatrix} \vect{\theta}^{(k+1)} &  \U^{(k+1)} \end{bmatrix} - \mathbf{V}^{(k+1)})$ 
		
		\ENDWHILE
		
	\end{algorithmic}
\end{algorithm}


\begin{restatable}{proposition}{prop1}
\label{theorem:scc-conv}(SCC convergence)
If $\curl$ is convex, Algorithm~\ref{alg:scc-alg} converges to a global solution. In addition, if $\curl$ is strictly convex, it converges to the unique global solution.
\end{restatable}

Proposition \ref{theorem:scc-conv} is an extension of Theorem 4 in \citet{wang2019integrative} and guarantees the convergence of multi-block ADMM using inexact sub-problem approximations.  
Similarly, to solve the supervised convex biclustering problem, we apply multi-block ADMM and provide Algorithm~\ref{alg:bicscc-alg} in Appendix \ref{appA}.

\subsection{Practical Issues}\label{practical}

In this section, we address some practical issues of applying our methods to real data. First, we show how to choose the regularization parameter $\lambda$. Then, we discuss the choice of weights and tuning parameters for the level of supervision. Moreover, we introduce an adaptive method to adjust for additional covariates.

\subsubsection{Choice of Regularization Parameter}

As mentioned, the tuning parameter $\lambda$ regulates both the number of clusters and the cluster assignments. The same type of procedures people use to choose regularization parameter $\lambda$ for other convex clustering methods work here. For example, \citet{wang2010consistent,fang2012selection} proposed stability selection based methods while \citet{chi2017convex} proposed hold-out validation.  In this paper, we suggest using stability selection when the number of clusters is not known. We find this approach works well in practice.

\subsubsection{Choice of Weights and Level of Supervision}\label{weights}

In practice, the choice of fusion weights has been shown to play an important role in computational efficiency and clustering quality. \citet{chi2015splitting,hocking2011clusterpath,chi2017convex} have shown that setting weights inversely proportional to the distances between two observations yields superior performance.  On the other hand, enforcing sparse weights reduces computational cost and improves clustering quality. Given these two, the most commonly used weights choice for convex clustering is $k$-nearest-neighbors method with a Gaussian kernel. 

The challenge here is that the unlabeled data $\X$ and supervising auxiliary variable $\y$ are measured from different sources and can be of different types; thus the Gaussian kernel with Euclidean distances is not an appropriate distance metric in this case. To measure the dissimilarity of two subjects measured in different data types, we adopt the Gower distance \citep{gower1971general}, which is a commonly used distance metric for mixed types of data and shown to obtain superior performance compared with other distance metrics \citep{wang2019integrative,ali2013k,hummel2017clustering}. The Gower distance between observation $i$ and $j$ can be defined as $g(\X_{i.},\X_{j.}) = \sum_{l=1}^{p} g_{ijl}/ p $ where $g_{ijl} =  \frac{ | \X_{il} - \X_{jl} |}{R_l}$ refers to the Gower distance between observation $i$ and $j$ for the $l^{\text{th}}$ feature  and $ R_l = \max_{i , j} | \X_{il} - \X_{jl} |$ is the range of the $l^{\text{th}}$ feature. 

Denote $\nu_{ij}^k$ as the indicator which equals 1 if observation $j$ is among observation $i$'s $k$ nearest neighbors or vice versa, and 0 otherwise. Let $\alpha$ be a non-negative tuning parameter bewtween 0 and 1. Then, our recommendation for weights is given by the following: 
$$w_{ij} = \nu_{ij}^k \exp{[-\phi(\:\:(1-\alpha) g(\mathbf{X}_{i\cdot},\mathbf{X}_{j\cdot}) + \alpha g(y_i , y_j)\:\:)]}.$$
The tuning parameter $\alpha$ suggests the level of supervision $\y$ gives to data $\X$. A larger $\alpha$ suggests putting more weight on the supervising auxiliary variables. In practice, we suggest choosing $\alpha = \frac{D_y} { D_y +  \| \mathbf{X}  - \bar{\mathbf{X}}  \|_F^2}$, where $D_y$ is the null deviance of supervising auxiliary variable$\y$, and hence $\alpha$ is the ratio of null deviances between two sources. If the clustering signal in the supervising auxiliary variable is weak, our choice of weight will down-weight this variable and vice versa. 
This weight scheme balances the contribution of $\X$ and $\y$. Yet, one may choose other weighting schemes based on the level of confidence in the supervising auxiliary variable as well.



In the presence of additional covariates, we can no longer calculate weights based on the distance between $y_i$ and $y_j$ as the supervising auxiliary variable $\y$ now contains the effect of additional covariates $\Z$. Recall the fusion penalty is measuring the dissimilarity between cluster centroids. To remove the effect of additional covariates in calculating weights, we suggest i) first estimating the effect of covariates $\hat \bbeta$ by fitting supervised convex clustering with weights not adjusted for covariates as usual and ii) then calculating weights based on distances between the supervising auxiliary variable centroids $\vect{\hat \theta}$ by removing the effect of covariates from the supervising auxiliary variable. This gives our adaptive supervised convex clustering with covariate-adjusted weights, as detailed in Algorithm \ref{alg:adaptive-covariate}. Our adaptive supervised convex clustering is similar to many adaptive approaches in the literature \citep{zou2006adaptive}.  

\begin{algorithm}[H]
	\caption{Adaptive SCC with covariate-adjusted weights}
	\label{alg:adaptive-covariate}
	\begin{algorithmic}
		\STATE {1. Fit SCC with $\bw$ and a sequence of $\gamma$;  Find $\gamma^{(k)}$ which gives desired number of clusters; Get the estimate  $\hat \bbeta^{(k)}$}. \\ 
	    \STATE {2. Update fusion weights: $\hat w_{ij} = \nu_{ij}^k \exp{[-\phi(\:\:(1-\hat \alpha) g(\mathbf{X}_{i\cdot},\mathbf{X}_{j\cdot}) + \hat \alpha g(\hat y_i , \hat y_j)\:\:)]}$}. Also update $\hat \alpha = \frac{ D_{\hat y}} {  D_{\hat y} +  \| \mathbf{X}  - \bar{\mathbf{X}}  \|_F^2}$, where $\hat y$ is the residual of supervising variable adjusted for the covariates.\\
	    \STATE {3. Fit  SCC with $\hat \bw$.}
	\end{algorithmic}
\end{algorithm}

To remove the effects of additional covariates on the supervising auxiliary variable, we use the property of the link function: $g(\mathbb E[\y|\mathbf{Z}]) = \vect{\theta} + \Z \bbeta$.  A good estimate of $\vect{\theta}$ would be removing the effect of covariate from the link function of $\y$, i.e., $ \vect{\hat \theta} = g(\y) - \Z \hat \bbeta$. Hence we can get estimated supervising auxiliary variable without the effect of additional covariates using inverse link function: $ \hat \y = g^{-1}(\vect{\hat \theta})$.

\section{Simulation Studies}\label{sim}
In this section, we evaluate the performance of supervised convex clustering and compare with existing methods. For all the simulations, we design challenging scenarios where clustering either $\X$ or $\y$ alone cannot lead to good clustering results. We will discuss the simulation setup in detail later.

We compare our supervised convex clustering method with convex clustering and hierarchical clustering using different distance metrics (Euclidean distances and Gower distances). It should be pointed out that there are several linkage options for hierarchical clustering and we only report the linkage with the best clustering performance. We use the adjusted Rand index \citep{hubert1985comparing} to evaluate the accuracy of clustering results.
The adjusted Rand index is a commonly used metric to measure the agreement between the estimated cluster label and the true underlying label.  A larger adjusted Rand index (close to 1) indicates a good resemblance between the estimated and true labels. For all methods, we assume that the oracle number of clusters is known for fair comparisons. 

We first consider the base simulation where the supervising auxiliary variable $\y$ is generated from the cluster centroid directly without additional covariates. In the base simulation, we study two designs of unlabeled data in which two different scenarios are considered. In addition to the base simulation, we consider other setups including varying dimensions, unequal group sizes and additional covariates.

The base simulation, as mentioned, assumes that the supervising auxiliary variable $\y$ is generated from the cluster centroid directly. For each simulation, the data set consists of $n=120$ observations and $p=30$ features with 3 clusters. Each cluster has an equal number of observations for the base simulation. The data is generated from the following model: $\X_{i.} \sim N(\mathbf \mu_k,\sigma^2 \textbf I_p)$, where $i \in G_k$, $k = 1,2,3$ ($G_k$ refers to the observation indices belonging to group $k$).  The supervising auxiliary variable, $y_i$, is generated from different distributions  with parameter $\mathbf \mu_k$ based on data type; the two sources have the shared group label which means $y_i \sim \phi(\mu_k)$, where $i \in G_k$, $k = 1,2,3$ and $\phi$ is a distribution function. We denote $\X_{G_k}$ and $\y_{G_k}$ as the data points and their corresponding supervising auxiliary variable that belong to group $k$.

We consider two designs of the unlabeled data $\X$: spherical (S) and half-moon (H). In terms of the half moon data, we consider the standard simulated data of three interlocking half moons as suggested by \cite{chi2015splitting} and \cite{wang2019integrative}. For each design, we consider two scenarios where none of the data sources lead to perfect clustering results. In the first scenario (S1 and H1), 
$\X_{G_1}$ and $\X_{G_3}$ overlap while $\X_{G_2}$ are separate from $\X_{G_1}$ and $\X_{G_3}$; $\y_{G_1}$ and $\y_{G_3}$ have two separate clusters while $\y_{G_2}$ are noisy and overlap with $\y_{G_1}$ and $\y_{G_3}$. 
In the second scenario (S2 and H2), $\X_{G_1}$, $\X_{G_2}$ and  $\X_{G_3}$ all overlap; $\y$ has three separate clusters with little overlapping. Yet, $\y$ is noisy and one cannot get perfect results by clustering $\y$ alone. Hence, for each of the four simulation scenarios above, we create a challenging problem where good clustering results cannot be achieved by clustering either $\X$ or $\y$ alone and we seek to get good clustering results by borrowing strength from two sources with our method. The detail of the exact parameters for simulating the unlabeled data and supervising auxiliary variables are given in Appendix \ref{appB}.

For the above simulations, we assume that the number of cluster centroids for both sources is the same. Yet, in the case of categorical supervising auxiliary variable, usually, the number of categories we observe in that variable is different from the number of true classes. Hence we consider the following additional simulations.  In additional simulation 1 (AS1), we assume the number of clusters of $\X$ is greater than number of categories in $\y$; in AS1, we consider both binary and categorical supervising auxiliary variables. In additional simulation 2 (AS2), we assume the number of categories in $\y$ is greater than number of clusters of $\X$; in AS2, we consider categorical supervising auxiliary variable.
    
From Table~\ref{scc-base}, we see that our supervised convex clustering outperforms existing methods for different types of supervising auxiliary variables by leveraging information from both sources. 
For hierarchical clustering on spherical data, different distance metrics might perform comparably well on different types of supervising auxiliary variable. For example, hierarchical clustering with Euclidean distances works well for Gaussian and count-valued supervising auxiliary variables while hierarchical clustering with Gower distances works well for binary and categorical supervising auxiliary variables. Yet, our SCC performs comparatively well in terms of the best hierarchical clustering method for all cases. For non-spherical data, our method performs significantly better than hierarchical clustering.

\begin{table}[!ht]
	\vskip 0.15in
	\begin{center}
\resizebox{\linewidth}{!}{%
  \begin{tabular}{lccccccc}
    \toprule
    \multirow{3}{*}{} &
      \multicolumn{4}{c}{Gaussian} &
      \multicolumn{3}{c}{Binary}  \\ 
     \cmidrule(lr){2-5}
     \cmidrule(lr){6-8} 
      & {S1} & {S2} & {H1} & {H2} & {S1} & {H1} & {AS1}        \\
     \midrule
    Hclust on X  & 0.71 (2.2e-2) & 0.64 (3.2e-2) & 0.20 (2.5e-2) & 0.19 (7.4e-3) &  0.69 (1.2e-2) & 0.34 (9.4e-2) & 0.48 (2.3e-2) \\
    Hclust on y   & 0.35 (1.4e-2) & 0.85 (4.0e-2) & 0.58 (2.2e-2) & 0.91 (1.3e-2) & 0.31 (6.8e-4)  & 0.35 (1.3e-3) & 0.64 (1.3e-2) \\
    Hclust on [X y]  & 0.93 (1.4e-2) & 0.87 (2.4e-2) & 0.58 (1.1e-1) & 0.39 (2.3e-2) & 0.74 (1.7e-2) & 0.40 (8.7e-2)& 0.44 (1.0e-2)\\
    Hclust on [X y] Gower  & 0.82 (4.5e-2) & 0.91 (1.9e-2) & 0.55 (2.2e-2) & 0.22 (1.1e-2) & 0.84 (2.1e-3) & 0.51 (5.6e-2) & 0.92 (5.2e-2)  \\
    Convex Clustering on X      &  0.56 (1.5e-3)  &  0.21 (6.3e-2)  &  0.62 (2.2e-2)  &  0.13 (1.4e-2)  &  0.56 (1.5e-3)    & 0.66 (1.1e-2)    &  0.44 (0.0e-0) \\
    Convex Clustering on [X y]  &  0.86 (6.6e-2)  &  0.58 (4.6e-2) &   0.99 (3.3e-3)   & 0.60 (6.5e-2)    &  0.56 (1.5e-3)    & 0.66 (1.2e-2)    &  0.44 (0.0e-0) \\
    SCC    & \textbf{0.96 (1.1e-2)} & \textbf{0.95 (9.5e-3)} & \textbf{1.00 (3.3e-3)} & \textbf{0.97 (1.0e-2)}  & \textbf{0.85 (4.6e-3)} & \textbf{0.80 (3.3e-2)} & \textbf{0.97 (1.1e-2)}   \\ 
    \bottomrule
  \end{tabular}
  }
	\end{center}
	\vskip -0.1in
	\vskip 0.15in
	\begin{center}
\resizebox{\linewidth}{!}{%
  \begin{tabular}{lcccccc}
    \toprule
    \multirow{1}{*}{} &
      \multicolumn{6}{c}{Categorical} \\
           \cmidrule(lr){2-7}
      & {S1} & {S2} & {H1} & {H2} & {AS1} & {AS2}    \\
      \midrule
    Hclust on X  &   0.71 (2.3e-2)  & 0.63 (2.0e-2)  & 0.34 (9.1e-2) & 0.20 (4.4e-3) &  0.51 (3.4e-3) & 0.51 (7.5e-3) \\
    Hclust on y   & 0.32 (5.2e-3) & 0.81 (1.6e-2)   & 0.43 (3.8e-3) & 0.81 (1.5e-2)  & 0.78 (4.3e-3)  & 0.51 (6.0e-4) \\
    Hclust on [X y]    & 0.75 (2.7e-2)  & 0.67 (2.7e-2)   & 0.34 (9.1e-2) & 0.20 (5.5e-3) & 0.52 (3.7e-17) & 0.51 (6.3e-3) \\
    Hclust on [X y] Gower  & 0.82 (5.3e-3) & 0.87 (1.2e-2)    & 0.56 (2.0e-3)  & 0.26 (1.1e-2)  & 0.35 (0.0e-0)  & 0.41 (3.0e-2) \\
    Convex Clustering on X        &  0.56 (1.2e-3)   &  0.14 (5.7e-2)    &  0.62 (2.4e-2)   & 0.15 (1.0e-2)   &  0.30 (8.6e-3)  &  0.78 (7.1e-2)    \\
    Convex Clustering on [X y]    &  0.56 (1.2e-3)   &  0.11 (5.7e-2)    &  0.70 (2.1e-2)   & 0.18 (1.1e-2)  &  0.46 (3.0e-2) &  0.90 (5.6e-2)     \\
    SCC   & \textbf{0.86 (2.0e-2)} & \textbf{0.89 (1.2e-2)} & \textbf{0.95 (3.3e-3)} & \textbf{0.83 (1.6e-2)} &   \textbf{0.85 (1.4e-2)} & \textbf{1.00 (0.0e-0)} \\
    \bottomrule
  \end{tabular}
  }
	\end{center}
	\vskip -0.1in
	\vskip 0.15in
	\begin{center}
\begin{adjustbox}{min height = 2\textheight, max width=1\textwidth}
  \begin{tabular}{lcccccccc}
    \toprule
    \multirow{3}{*}{} &
      \multicolumn{4}{c}{Count} &
      \multicolumn{4}{c}{Survival}  \\
    \cmidrule(lr){2-5}
     \cmidrule(lr){6-9}
      & {S1} & {S2} & {H1} & {H2} & {S1} & {S2} & {H1} & {H2}  \\
      \midrule
    Hclust on X  & 0.73 (2.4e-2) & 0.65 (1.8e-2) & 0.40 (8.2e-2) & 0.18 (7.1e-3)   &  0.73 (2.0e-2) & 0.65 (2.0e-2) & 0.36 (8.7e-2) & 0.27 (4.0e-2)     \\
    Hclust on y   &0.37 (7.5e-3)  & 0.75 (4.3e-2) & 0.43 (2.3e-2)& 0.82 (1.1e-2)  & 0.10 (2.2e-2)  & 0.09 (1.3e-3) & 0.14 (3.3e-2)& 0.12 (1.5e-2)      \\ 
    Hclust on [X y]  &  0.88 (2.8e-2) & 0.94 (1.3e-2)& 0.55 (4.6e-2) & 0.94 (9.0e-3)& 0.73 (2.0e-2) & 0.65 (2.3e-2)& 0.34 (9.4e-2) & 0.28 (2.9e-2)   \\ 
    Hclust on [X y] Gower  & 0.85 (4.1e-2) & 0.91 (1.3e-2) & 0.57 (2.9e-2) & 0.20 (1.1e-2)  & 0.59 (2.6e-2) & 0.55 (9.9e-2) & 0.59 (1.9e-2) & 0.25 (1.1e-2) \\
    Convex Clustering on X       &  0.56 (1.5e-3)  &  0.03 (1.6e-2)    &  0.64 (1.3e-2)   &  0.13 (9.5e-3)   &  0.56 (2.1e-3)   &  0.01 (1.1e-3)    &  0.68 (1.2e-2)   &  0.20 (1.8e-2)  \\
    Convex Clustering on [X y]   &  0.82 (5.5e-2)  &  0.80 (6.5e-2)    &  0.83 (5.9e-2)    & 0.96 (6.0e-3)    & 0.56 (1.5e-3)    & 0.01 (1.1e-3)    &  0.67 (1.3e-2)   &  0.11 (1.5e-2)    \\ 
    SCC    & \textbf{0.94 (1.8e-2)} & \textbf{0.96 (6.7e-3)} & \textbf{0.97 (2.6e-2)} & \textbf{0.98 (4.5e-3)}  & \textbf{0.87 (1.6e-2)} & \textbf{0.81 (2.4e-2)} &\textbf{ 0.92 (5.0e-2)} &\textbf{ 0.87 (3.6e-2)} \\ 
    \bottomrule
  \end{tabular}
  \end{adjustbox}
    \caption{{\small \textit{Comparisons of adjusted Rand index for supervised convex clustering and existing methods; Base simulation for Gaussian, binary, categorical, count-valued and censored survival supervising auxiliary variables.}}}
    \label{scc-base}
	\end{center}
	\vskip -0.1in
\end{table}



For the rest of this section, we consider different setups from the base simulation and verify that our method could still perform well in these settings. We slightly change the setup of scenario 1 (S1) in the base simulation described above  
with Gaussian supervising auxiliary variable. Specifically, we vary the number of features and group sizes one at a time while keeping the rest of the setup the same. We then consider the case when the supervising auxiliary variable is affected by additional covariates. Finally, we examine the performance of supervised convex biclustering.

First, we change the number of features of $\X$ from 30 to 50 and 100 respectively. To increase the difficulty of the simulation, we increase the within-cluster variance $\sigma$ in the simulation setup. From Table~\ref{scc-add}, we see that our supervised convex clustering method still performs comparably well as the number of features increases by leveraging the information from both two sources. In contrast, existing methods do not perform as well as in Table~\ref{scc-base} since $\X$ now contains more clustering information with increased dimension and dominates the clustering results for existing methods.   Moreover, we evaluate the performance of our method on unequal group sizes with $n_1 = 80$, $n_2 = 10$ and $n_3 = 30$. Table~\ref{scc-add} suggests that our method also performs better than existing methods in this setup.

\begin{table}[!ht]
	\vskip 0.15in
	\begin{center}
  \begin{tabular}{lccc}
    \toprule
    \multirow{1}{*}{}       & {$p = 50$} & {$p = 100$} & Unequal group sizes   \\
      \midrule
    Hclust on X  &   0.65 (2.0e-2)  & 0.76 (1.8e-2)  & 0.64 (3.9e-2) \\
    Hclust on y   & 0.39 (4.3e-2) & 0.39 (4.3e-2)   & 0.68 (4.4e-2) \\
    Hclust on [X y]    & 0.84 (2.4e-2)  & 0.85 (1.4e-2)   & 0.91 (2.5e-2)\\
    Hclust on [X y] Gower  & 0.72 (4.8e-2) & 0.79 (4.0e-2)    & 0.83 (4.0e-2)  \\
    Convex Clustering on X       &  0.56 (1.5e-3)    &  0.56 (1.4e-3)    &  0.32 (2.8e-3) \\
    Convex Clustering on [X y]   &  0.60 (4.2e-2)    &  0.56 (1.4e-3)    &  0.64 (1.1e-1)   \\
    SCC   &  \textbf{0.94 (1.6e-2)} &  \textbf{0.93 (1.9e-2)}  & \textbf{0.97 (5.8e-3)}  \\
    \bottomrule
  \end{tabular}
  \caption{{\small \textit{Comparisons of adjusted Rand index for supervised convex clustering and existing methods; Additional simulation for Gaussian supervising auxiliary variables; the data is simulated from the same setup as S1 for Gaussian supervising auxiliary variable in the base simulation, but with different number of features and unequal group sizes.}}}
    \label{scc-add}
	\end{center}
	\vskip -0.1in
\end{table}



Next, in the following simulation, we examine the performance of our supervising convex clustering when the supervising auxiliary variable is affected by additional covariates. We use our adaptive supervised convex clustering approach proposed in 
Section~\ref{weights}  to adjust for these additional covariates. 

The data $\X$ is generated from the same distribution as in scenario 1 (S1) of the base simulation described above. 
Still, we consider different types of supervising auxiliary variable. Yet, the supervising auxiliary variable $y_i$ is now simulated from  $y_i \sim \phi(\mu_k + \Z_i^T \bbeta)$ where $\Z_i \in \mathbb R^{10} \sim N(\textbf 0, \textbf I_{10})$ and $\beta_j \sim N(\pm 3,1)$; $\mu_k$ is generated similarly in Scenario 1 of the base simulation. We set the number of features for the additional covariates to be 10.

Table~\ref{scc-covariate} shows that our adaptive supervised convex clustering performs the best by removing the effects of additional covariates from supervising auxiliary variable; hence our method clusters objects based on the exact centroids which form the groups. On the other hand, existing methods all do poorly as they perform clustering based on the supervising auxiliary variable $\y$ which is affected by additional covariates.

\begin{table}[!ht]
	\vskip 0.15in
	\begin{center}
\resizebox{\linewidth}{!}{%
  \begin{tabular}{lccccc}
    \toprule
    \multirow{1}{*}{}       & {Gaussian} & {Binary} & {Categorical} & {Count} & {Survival}    \\
      \midrule
    Hclust on X  &    0.68 (2.1e-2)  & 0.73 (3.4e-2)   & 0.70 (2.8e-2) &  0.69 (2.6e-2)   & 0.73 (2.1e-2) \\
    Hclust on y   &  0.04 (1.0e-2) & 0.22 (2.6e-2) & 0.17 (1.8e-2) & 0.07 (7.9e-3)   & 0.05 (1.8e-2) \\
    Hclust on [X y]    &  0.23 (5.8e-2) & 0.76 (3.0e-2)   & 0.70 (2.0e-2)  & 0.56 (2.7e-2)   & 0.74 (2.0e-2) \\
    Hclust on [X y] Gower  & 0.57 (2.8e-2) & 0.75 (2.9e-2)    & 0.58 (3.3e-2) & 0.66 (5.2e-2)  & 0.59 (1.5e-2) \\
    Convex Clustering on X          & 0.56 (1.2e-3)    &   0.56 (1.5e-3)    &  0.56 (3.0e-3)    &   0.56 (1.5e-3)  &  0.56 (1.4e-3) \\
    Convex Clustering on [X y]      & 0.40 (6.4e-2)    &  0.56 (1.5e-3)     &  0.55 (4.9e-3)    &   0.58 (1.2e-2)  &  0.56 (1.4e-3)  \\
    Adaptive SCC   & \textbf{0.99 (1.4e-2)} & \textbf{0.85 (2.7e-2)}  & \textbf{0.94 (1.1e-2)} & \textbf{0.89 (2.2e-2)} & \textbf{0.83 (3.2e-2)}\\
    \bottomrule
  \end{tabular}
  }
  \caption{{\small \textit{Comparisons of adjusted Rand index for supervised convex clustering and existing methods; Supervising auxiliary variable affected by additional covariates; the unlabeled data is simulated from the same setup as S1 in the base simulation, but the supervising auxiliary variables are simulated from different centroids affected by covariates.}}}
    \label{scc-covariate}
	\end{center}
	\vskip -0.1in
\end{table}



Finally, in this simulation setup, we evaluate the performance of our supervised convex biclustering method on scenario 1 (S1) in the base simulation described above 
with Gaussian supervising auxiliary variable.

Table~\ref{scc-supbicvx} suggests that our supervised convex biclustering method performs as well as supervised convex clustering method in Table~\ref{scc-base}. Yet, our supervised convex biclustering method groups similar features simultaneously and identifies checkerboard-like patterns.

\begin{table}[!ht]
	\vskip 0.05in
	\begin{center}
		\begin{small}
				\begin{tabular}{lcccr}
					\toprule
					Method &  Adjusted Rand Index \\
					\midrule
					Hclust on X   & 0.69 (2.6e-2) \\
					Hclust on y  &   0.39 (1.3e-2) \\
					Hclust on [X y]  &   0.95 (9.8e-3) \\
					Hclust on [X y] with Gower & 0.84 (4.8e-2) \\
				Convex Biclustering on X          & 0.67 (3.7e-2)    \\
                Convex Biclustering on [X y]      & 0.89 (5.6e-2)   \\
					Supervised Convex Biclustering  &  \textbf{0.98 (6.5e-3)} \\
					\bottomrule
				\end{tabular}
	\caption{{\small \textit{Comparisons of adjusted Rand index for supervised convex biclustering and existing methods; the unlabeled data and supervising auxiliary variable are simulated from the same setup as S1 for Gaussian supervising auxiliary variable in the base simulation.}}}
	\label{scc-supbicvx}
		\end{small}
	\end{center}
	\vskip -0.1in
\end{table}

Overall, we demonstrate the strong empirical performance of our supervised convex clustering method which includes the information of both the unlabeled data $\X$ and supervising auxiliary variable $\y$ to get better clustering results.

\section{Case Study: Discovering New Subtypes of Alzheimer's Disease}\label{case_study}

An important application of our proposed method is in clinical genomics, where the objective is to elucidate the genetic basis of diseases and to find potential biomarkers for developing personalized treatments. An important aspect of personalized treatments is to identify groups of subjects with similar genetic profiles and similar clinical outcomes so that personalized medicine targeting specific gene groups can be developed. For example, breast cancer patients with different genomic subtypes now receive different sets of treatments. We would like to investigate the clinical genomics of AD to better study the genetic basis of AD. Alzheimer's Disease (AD) is a debilitating brain disorder that irreversibly damages cognitive skills. However, there is a large amount of heterogeneity in cognition of older adults and little is known about the underlying genetic mechanisms that cause AD apart from a handful of genes. In this case study, we apply our SCC method to find biologically meaningful group structures among both subjects and genomic profiles for AD by jointly analyzing both clinical measurements and gene expression via RNASeq acquired from the Religious Orders Study Memory and Aging Project (ROSMAP) Study \citep{bennett2018rosmap}. 

For our analysis, to start with, we take the clinically measured global cognition score as the noisy supervising auxiliary variable. Global cognition score, a summary measure of cognition proximal to death, is computed by averaging nineteen clinical cognitive tests conducted during a subject's last clinical visit \citep{bennett2018rosmap}. A higher value for global cognition score indicates relatively healthier cognitive abilities. The ROSMAP data consists of $507$ subjects with a complete recording of $41,809$ RNASeq genes. First, we log-transform the RNASeq counts, which is commonly done in many RNASeq analyses. After that, we remove undesirable batch effects from RNASeq using the ComBat technique \citep{johnson2007combat}. To reduce the number of genes to a more manageable size, we take the top $20,000$ RNASeq genes with the highest variance and subsequently keep the top $600$ genes that are most associated with cognition score via univariate filtering.

\begin{figure}[H]
\centering
\includegraphics[width=1\textwidth]{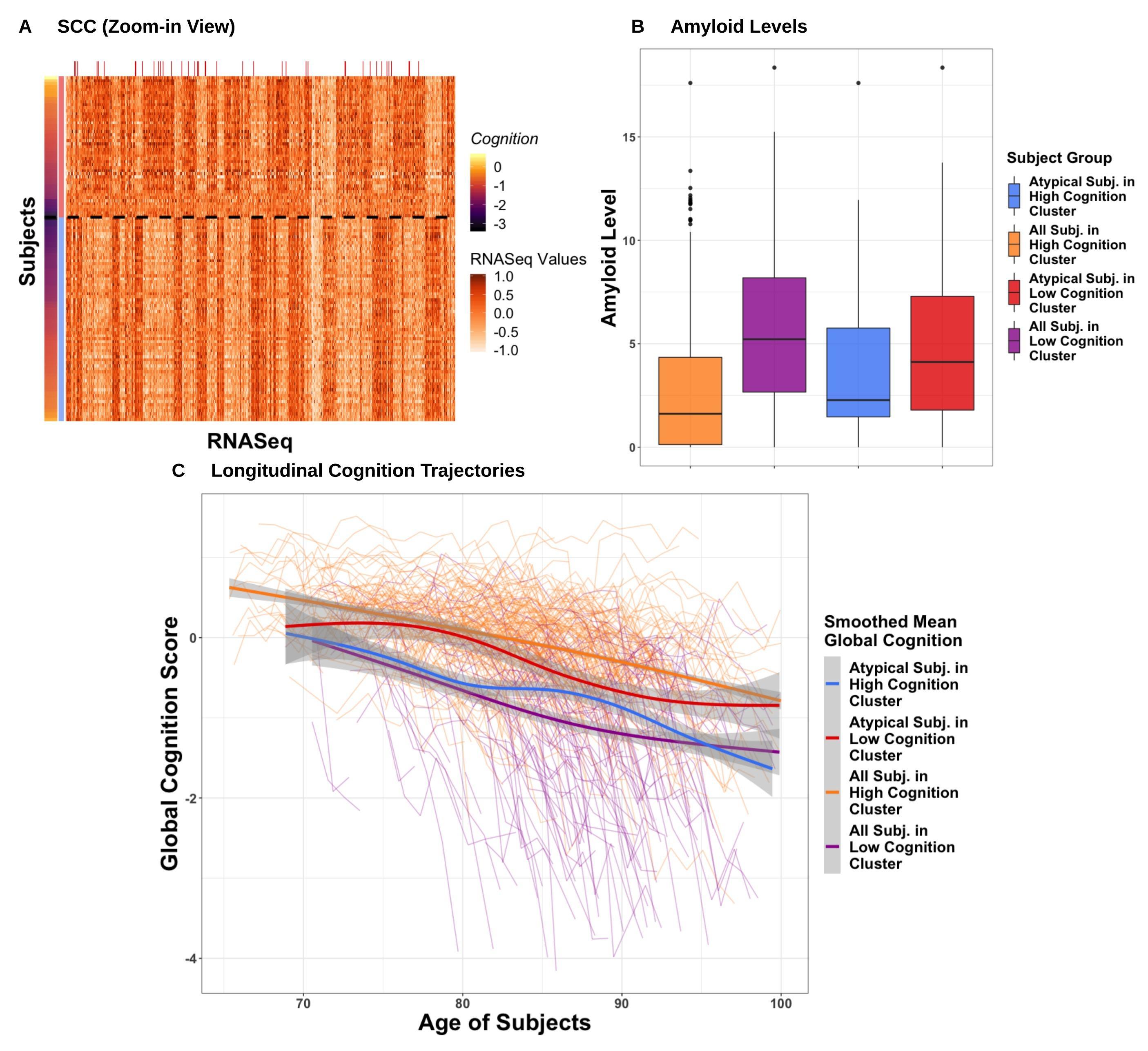}
\caption{{\small \textit{(A) Zoom-in view of the RNASeq heatmap produced by SCC. Atypical subjects with unusually high cognition in the low cognition cluster are highlighted with a red bar while atypical subjects with much poorer cognition in the high cognition cluster are highlighted with a blue bar on the left. (B) We plot the longitudinal trajectories of cognition score of all subjects with bolded lines representing smoothed mean cognition of the various subgroups identified by SCC. (C) Boxplots of amyloid plaque levels of the various subgroups identified by SCC. Collectively, the figures provide evidence that the atypical subjects in the low cognition cluster could be Cognitive Resilient (CR) while the atypical subjects in the high cognition cluster may have non-AD related dementia.}}}
\label{fig:zoominheatmap}
\end{figure}

Our goal is to identify both scientifically meaningful group structures among subjects and potential genetic biomarkers for AD that may be hidden in completely supervised/unsupervised analyses of data by leveraging information from both clinical global cognition score as supervising auxiliary variable and unlabeled RNASeq gene expression data. To this end, we run our SCC-biclustering method with adjustment for age at death, which simultaneously estimates group structures among subjects and RNASeq genes from the preprocessed ROSMAP data. We include adjustment for age to account for its effects on the cognition score because cognitive skills are expected to decline as an individual ages, as shown in Fig \ref{fig:zoominheatmap}C.

\begin{figure}[H]
\centering
\includegraphics[width=1\textwidth,height = 8cm]{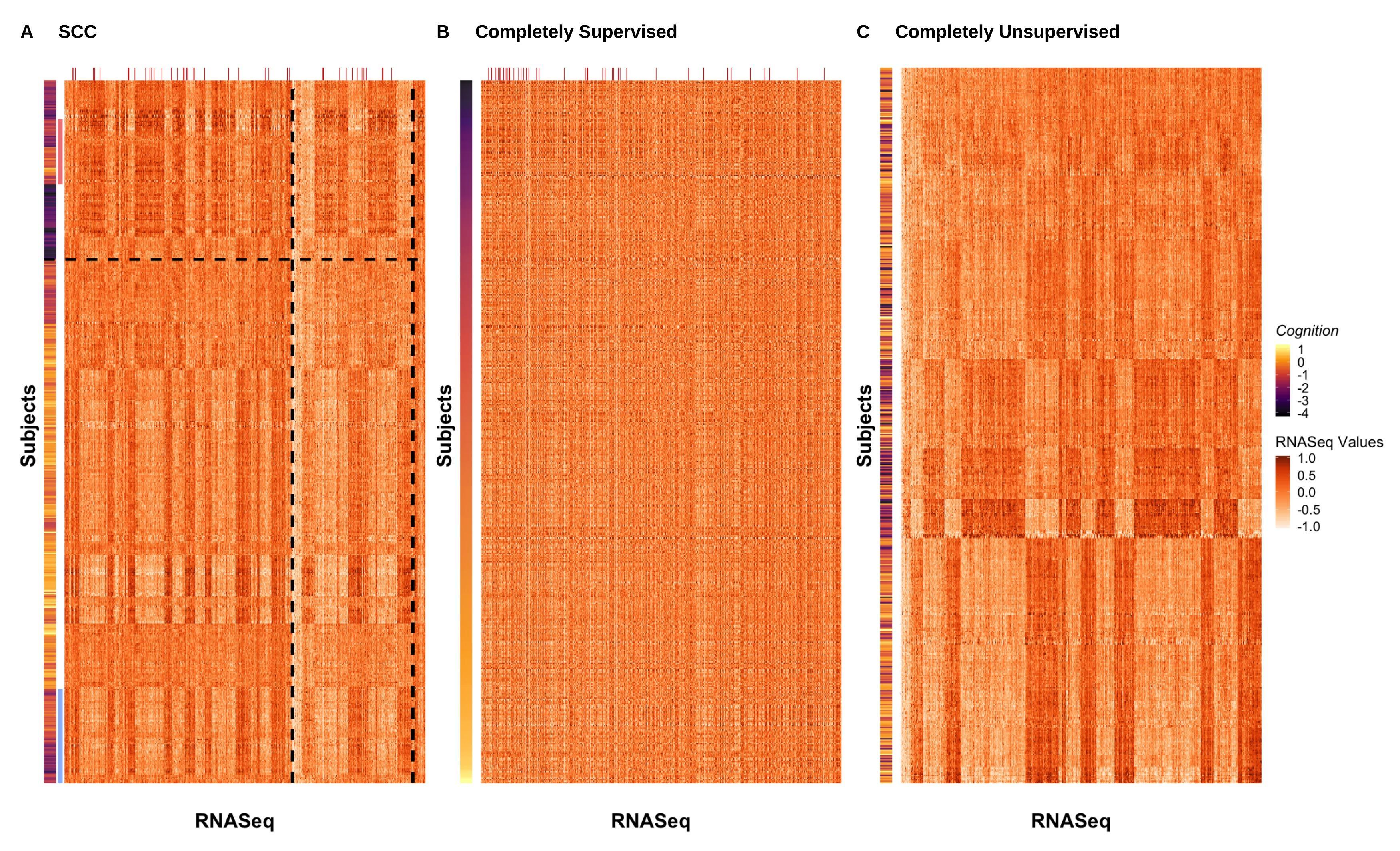}
\caption{{\small \textit{(A) The subjects and genes are ordered according to the cluster assignment estimated by SCC-biclustering (adjusting for age at death). Cluster boundaries are indicated with black dashed lines. Atypical subjects in the high cognition cluster (low cognition cluster) are hightlighted with blue (red) on the left. Top 40 DEGs whose median expression levels are significantly different across the two SCC clusters at FWER of level 0.05 are highlighted with red bars on top. (B) The subjects are ordered in ascending order of cognition score and the genes are ordered in ascending order of p-values obtained from univariate association tests of each gene with cognition score adjusting for age. The rank of the DEGs found by SCC in terms of the univariate association p-values are indicated with red bars. (C) The subjects and genes are ordered according to dendrograms obtained from hierarchical biclustering on RNASeq data alone.}}}
\label{fig:compareheatmap}
\end{figure}

The resulting heatmap of RNASeq profiles is displayed in Fig \ref{fig:compareheatmap}A, where we order the subjects (rows) and genes (columns) according to the cluster assignment estimated by our SCC-biclustering with black dashed lines indicating cluster boundaries. In addition, the corresponding global cognition score is displayed on the left side of the heatmap. For comparison, we also show a completely supervised approach that treats the global cognition score as the true response variable. The heatmap generated by this completely supervised approach is shown in Fig \ref{fig:compareheatmap}B, where the subjects are ordered by the ascending global cognition score (from top to bottom) while the genes are ordered in ascending order of p-values obtained from univariate association test of each gene with the global cognition score after adjusting for age. Finally in Fig \ref{fig:compareheatmap}C, subjects and genes are ordered according to the cluster heatmap obtained from completely unsupervised hierarchical biclustering on unlabeled RNASeq data with Euclidean distance metric and Ward linkage, as shown in Fig \ref{fig:compareheatmap}C.

Our results reveal, as shown in Fig \ref{fig:compareheatmap}A, the top cluster above the horizontal dashed line consists mostly of subjects with relatively low cognition scores whereas the bottom cluster is made up of subjects with generally higher cognition scores. For simplicity, we call the top cluster (bottom cluster) in Fig \ref{fig:compareheatmap}A  the ``low cognition cluster" (``high cognition cluster") thereafter. A quick examination of RNASeq gene signatures across the low and high cognition clusters obtained from our SCC reveals clear differences in expression levels of many RNASeq genes, indicating possible genetic biomarkers responsible for influencing cognitive decline and onset of AD. On the other hand, even though completely unsupervised clustering of RNASeq data results in a heatmap (Fig \ref{fig:compareheatmap}C) with seemingly distinct genetic profiles across clusters, these estimated clusters are much less scientifically interpretable as each cluster contains subjects across the entire spectrum of global cognition score. Furthermore, the heatmap produced by the fully supervised approach in Fig \ref{fig:compareheatmap}B lacks almost any distinguishable patterns that might reveal genetic differences between subjects with higher cognition and subjects with lower cognition. To briefly summarize, our SCC method leverages information from both unlabeled RNASeq gene expression data and noisy supervising auxiliary variable, global cognition score, to recover more interpretable clusters of subjects and gene signatures, in contrast to either fully supervised or unsupervised methods which only estimate clusters using one data source.

To study the scientific validity of the clusters discovered by our SCC method, we focus on analyzing the heterogeneity among subjects and RNASeq genes discovered by SCC (Fig \ref{fig:compareheatmap}A). First and foremost, many RNASeq genes appear to be upregulated (downregulated) for subjects in the low cognition cluster whereas these same genes seem to be downregulated (upregulated) for individuals in the high cognition cluster. To identify potential genetic biomarkers that might account for the differences in global cognition across the low and high cognition clusters, we extract the top $40$ differentially expressed genes (DEGs) across the two clusters, which are highlighted with short red bars on top of the heatmap in Fig \ref{fig:compareheatmap}A. We define a gene to be differentially expressed if its median expression levels are significantly different across the two SCC-estimated clusters according to a Wilcoxon rank sum test with Familywise Error Rate (FWER) controlled at level $0.05$. 

\begin{figure}[H]
\centering
\includegraphics[height  = 7.5cm, width=0.6\textwidth]{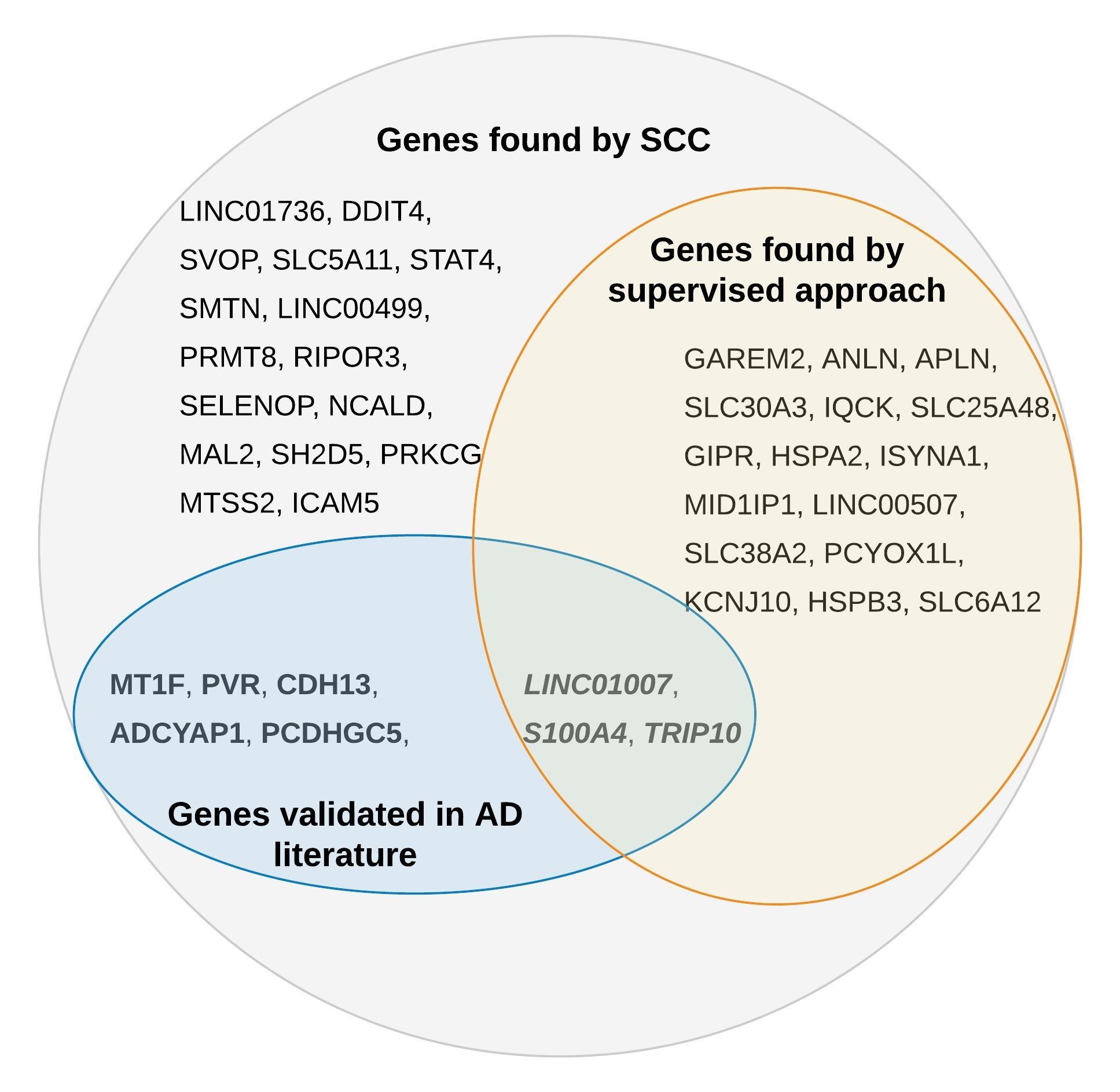}
\caption{{\small \textit{We plot a Venn diagram to show relations between top 40 DEGs discovered by our SCC and genes found by other methods. Gray circle: the top 40 DEGs found by SCC. Orange circle: intersection between DEGs found by SCC and top 100 genes that are significantly, univariately associated with cognition score. Blue circle: Genes that are found to be related to cognitive decline/AD pathology in the biological literature.}}}
\label{fig:venndiagram}
\end{figure}

The DEGs found by SCC are summarized in the gray circle of the Venn diagram shown in Fig \ref{fig:venndiagram}. For comparison, we also select the top $100$ RNASeq genes with the smallest p-values from the completely supervised univariate association tests with the global cognition score. The intersection between DEGs found by our SCC and genes that are significantly associated with cognition score are displayed in the orange circle in Fig \ref{fig:venndiagram}. After conducting a literature search on the top DEGs discovered by our SCC, we found evidence in the AD literature, which links at least eight of these DEGs to cognitive decline and/or AD pathology in AD patients \citep{liu2018gene,han2014gene, carter2017gene,gomez2010gene, pcdhgc5citation, linc01007, s100a4, trip10}. These eight DEGs are shown in the blue circle in Fig \ref{fig:venndiagram}. While this is only a preliminary investigation into the improved scientific interpretation and validity of clusters obtained from SCC, successfully identifying DEGs that have been validated in the AD literature is encouraging evidence. Additionally, the other 16 DEGs found by SCC, which could be missed by the supervised approach, may point to candidates for future studies of genetic basis for AD. Overall, this indicates that our SCC method yields results that are very distinct from other supervised learning approaches.

Beyond the discovery of potential biomarkers for possibly elucidating genetic mechanisms of AD, it is also scientifically interesting to examine the heterogeneity among subjects uncovered by SCC. In particular, even though the subjects in the low cognition cluster have very similar RNASeq expression patterns and overall lower cognition, it is worth noting that a small group of 48 atypical subjects, as highlighted with the red bar in Fig \ref{fig:compareheatmap}A, have unusually high global cognition. For simplicity, we refer to these 48 subjects as ``atypical subjects in the low cognition cluster''. Similarly, in the high cognition cluster, a small group of 68 subjects, as indicated by the blue bar in Fig \ref{fig:compareheatmap}A, have noticeably poorer global cognition than the rest of the cluster. We refer to these 68 subjects as ``atypical subjects in the high cognition cluster''.

To better understand this heterogeneity uncovered by SCC, we zoom into the RNASeq profiles of these atypical subjects, as shown in Fig \ref{fig:zoominheatmap}A. Surprisingly, although the two atypical subgroups have distinctly different RNASeq gene signatures, the median global cognition scores of these two subgroups are not significantly different according to a two-sided Wilcoxon rank sum test ($\textit{p-value}=0.068$). In Fig \ref{fig:zoominheatmap}C, we visualize the longitudinal trajectories of global cognition score of all $507$ subjects included in our case study with the bolded lines indicating smoothed mean global cognition of the various subgroups of subjects found by SCC. It is particularly interesting to note that even though atypical subjects in the low cognition clusters have very similar gene signatures to the rest of the low cognition cluster, which might be indicative of AD pathology, its mean longitudinal cognition (red curve in Fig \ref{fig:zoominheatmap}C) declines at a much slower rate and ends up with healthier cognition before death as compared to the mean cognition of the entire low cognition cluster (purple curve in Fig \ref{fig:zoominheatmap}C). Such seemingly contradictory observation hints at the possibility that these atypical subjects in the low cognition cluster might possess certain degrees of so-called Cognitive Resilience (CR), which is a phenomenon where healthy cognition can exist despite extensive AD-related pathology \citep{negash2011cognition, stern2012cognitive}. Previous scientific studies have also observed that individuals with higher CR experience a slower rate of cognitive decline over time \citep{yu2015residual}.

To further substantiate this finding, we examine levels of amyloid plaques, one of the hallmarks of AD brain pathology \citep{takahashi2017amyloid}, of the various subgroups identified by our SCC (Fig \ref{fig:zoominheatmap}B). Overall, the low cognition cluster (purple box in Fig \ref{fig:zoominheatmap}B) has significantly higher median amyloid level than the high cognition cluster (yellow box in Fig \ref{fig:zoominheatmap}B) according to one-sided Wilcoxon rank sum test ($\textit{p-value}=2.9\times 10^{-15}$). In the meantime, the mean cognition trajectory of the overall low cognition cluster is well below that of the high cognition cluster, likely due to AD-related cognitive decline. On the other hand, despite significantly higher median amyloid level of these atypical subjects in the low cognition cluster as compared to the high cognition cluster ($\textit{p-value}=3.3\times 10^{-5}$), the mean cognition trajectory of these atypical subjects in the low cognition cluster is fairly close to that of the high cognition cluster (yellow curve in Fig \ref{fig:zoominheatmap}C). In other words, the atypical subjects in the low cognition cluster found by SCC manage to maintain cognitive abilities on par with the relatively healthy high cognition cluster, even though these atypical subjects also possess high amyloid levels indicative of extensive AD brain pathology. Discovery of such atypical subject groups from the ROSMAP data provides new potential avenues for further scientific studies to better understand the genetic mechanisms responsible for the development of Cognitive Resilience, conferring potential clinical utility to our SCC method in clinical genomics. 

In addition to discovering subjects with high CR, our SCC method manages to identify subjects with dementia caused by conditions other than ``gold standard" AD pathology. Specifically, the atypical in the high cognition cluster appear to be free of RNASeq signatures that is common among subjects in the low cognition cluster (Fig \ref{fig:zoominheatmap}A). Also, median amyloid level of these atypical subjects in the high cognition cluster is significantly lower than that of the atypical subjects in the low cognition cluster ($\textit{p-value}=0.036$), as shown in Fig \ref{fig:zoominheatmap}B. Nonetheless, the atypical subjects in the high cognition cluster appear to experience a much steeper drop in mean cognition over time than the overall high cognition cluster as well as the possibly Cognitive Resilient subgroup, although aforementioned evidences suggest the atypical subjects in the high cognition cluster probably do not possess AD-related pathology. Further analyses reveal that lewy bodies are present in $19\%$ of the atypical subjects in the high cognition cluster while microinfarcts are present in another $43\%$ of these atypical subjects, both of which have been identified to be possible non-AD causes of dementia in previous studies \citep{mckeith1996consensus, arvanitakis2011microinfarct}. 


\begin{figure}[H]
\centering
\includegraphics[width=1\textwidth]{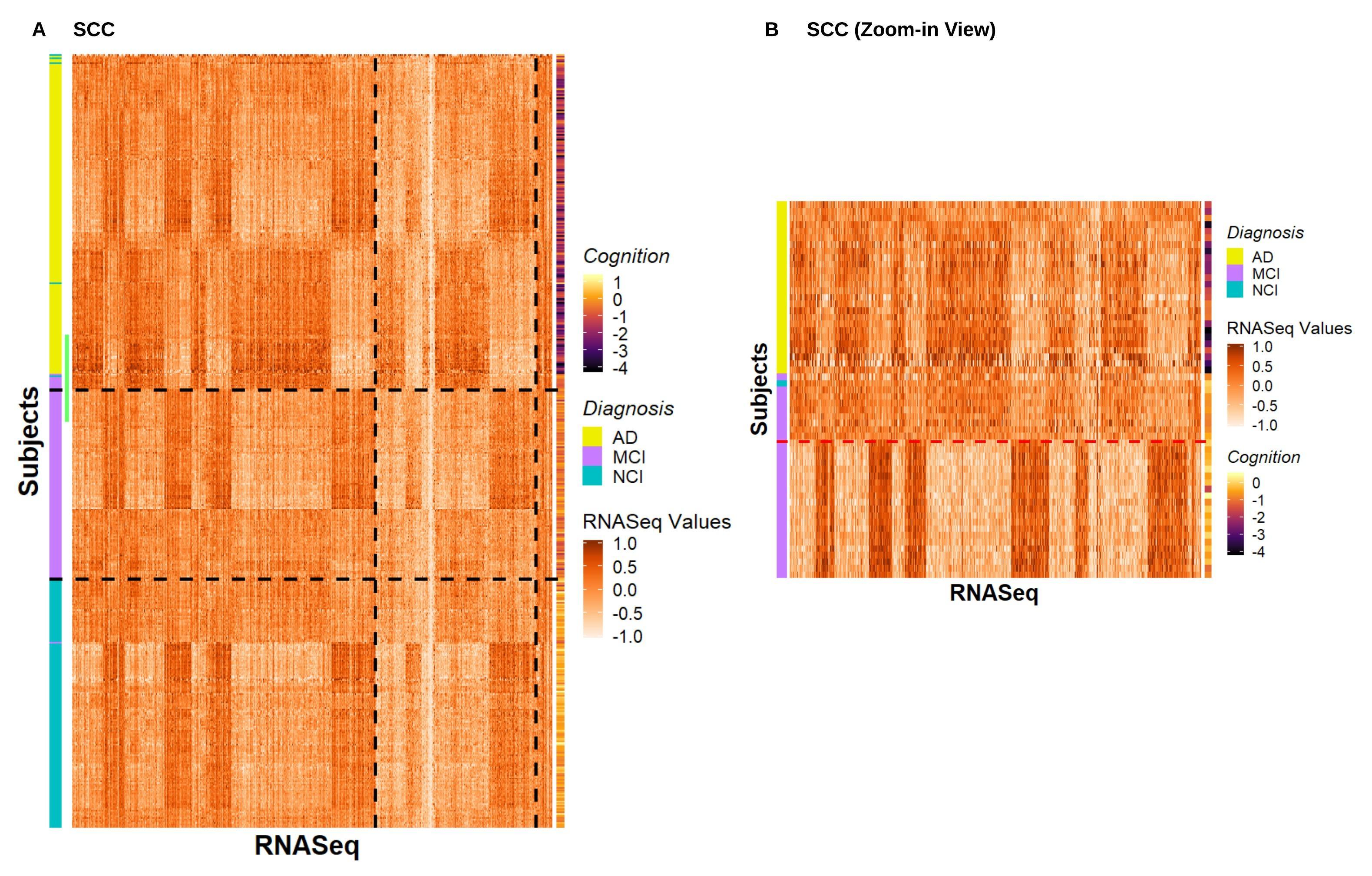}
\caption{{\small \textit{(A) The results of SCC where the subjects and genes are ordered according to the cluster assignment estimated by SCC-biclustering with clinician's diagnosis as supervising auxiliary variable (adjusting for age at death). Cluster boundaries are indicated with black dashed lines. Zoom-in subjects in (B) are hightlighted with green. Cognition scores are plotted on the right for reference. (B) A zoom-in plot of the heatmap reveals that several MCI subjects have gene expression profiles that are more similar to AD subjects.} }} 
\label{fig:scc-cogdx}
\end{figure}

Additionally, we also apply our SCC method to the ROSMAP data using the clinician's diagnosis as the supervising auxiliary variable. Clinician's diagnosis, a summary diagnostic opinion rendered by a neurologist prior to a patient's death, is a categorical variable with three levels - no cognitive impairment (NCI), mild cognitive impairment (MCI), and Alzheimer’s Disease (AD). Due to the large amount of heterogeneity in cognitive decline, there are no definitive standards to diagnose AD subtypes prior to death without postmortem pathology data. Therefore, clinician's diagnosis can be subjective and prone to judgement errors. We expect these diagnostic opinions to be noisy and can not fully trust them. Here, we would like to examine whether we can use genomics data to help improve these diagnostic opinions with our SCC method. Again, we apply our SCC-biclustering method with adjustment for age at death to simultaneously find group structures among subjects and RNASeq genes. Fig~\ref{fig:scc-cogdx}A shows the heatmap of RNASeq profiles where the subjects and genes are ordered according to the cluster assignment estimated by SCC-biclustering with clinician's diagnosis displayed on the left side. Overall, we see that the genomics can help us differentiate AD subtypes fairly well. Interestingly, by borrowing strength from signals in both the supervising auxiliary variable and unlabeled RNASeq data, our SCC method deems that a handful of MCI subjects should be grouped together with the majority of the AD subjects. This might first come as a surprise, but a close examination of RNASeq expressions of the aforementioned MCI subjects does reveal that the gene signatures of these MCI subjects above the dashed line in Fig \ref{fig:scc-cogdx}B indeed resemble those of AD subjects more than expressions of the rest of the MCI cluster. 
Hence our method uncovers joint group structure and identifies subjects whose diagnoses might need to be re-assessed. 



\section{Discussion}

In this paper, we develop a novel supervised convex clustering method that leverages the information from both supervising auxiliary variables and unlabeled data. Our method, in contrast to existing semi-supervised clustering approaches, is the first one to directly use outcome of interest to help cluster unlabeled data. In particular, our SCC borrows strength from both information sources and yields more scientifically interpretable group structures that may be hidden in completely unsupervised analyses of data.  

This paper mainly addresses the methodological development for supervised convex clustering but there are many possible open areas for future research. One potentially interesting area of future work may be to investigate supervised convex clustering with missing data or with missing supervising auxiliary variables. Handling missing data may be more amenable for our convex clustering based approach where \citet{chi2017convex} have developed extensions for missing data, than for other clustering techniques.  Another extension, based on the recent paper of \citet{wang2019integrative}, could be supervised convex clustering with data integration, where multiple sources of data or supervising auxiliary variables are observed.  Additionally \citet{wang2018sparse} and \citet{wang2019integrative} recently proposed to perform feature selection and convex clustering simultaneously, another extension that could be incorporated into our supervised convex clustering framework.  This paper focuses on methodological development, but we expect our approach to inherit many desirable theoretical properties of convex clustering and plan to investigate this in future work.  Finally, \citet{weylandt2019dynamic} recently proposed fast algorithms and visualization tools both static, dendrograms, and dynamic, clustering path plots, of the convex clustering solution. Their theoretical assumptions should apply in our supervised convex clustering setting and thus allow us to use dendrograms to additionally aid in visualizing our results.

One question that is worth further investigating is, whether we should use supervised convex clustering, and, how practitioners can tell whether the supervising auxiliary variable is useful for finding group structures. Further research could investigate when to apply supervised convex clustering and how much supervision is warranted for given problems.  We suggest a data-driven approach to determine the amount of supervision using the relative deviance in the two data sources. But one might adopt some other approaches, such as learning the amount of supervision from the data.

We apply our method to a high-dimensional genomics case study. Yet, our approach may find applications in a variety of fields such as electronic health records, online market segmentation, and text mining, among the many other clustering applications.  For example, in online market segmentation, some additional information on the users and the items are typically available, such as previous purchasing history, demographics, and social media usage, among others. We might use this meta information as supervising auxiliary variables to help understand joint group structures. To summarize, we develop a novel, unified approach to an interesting but challenging problem that leads to more scientifically interpretable clustering results and opens many avenues for future research.

\section*{Acknowledgements}

The authors acknowledge support from NSF DMS-1554821, and NSF NeuroNex-1707400. The  authors  thank  Dr. Joshua  Shulman for discussions on the ROSMAP data and acknowledge support from NIHP30AG10161,  RF1AG15819,  R01AG17917,  and R01AG36042 for this data.

\newpage
\begin{appendix}

\begin{center}
{\bf \LARGE Supervised Convex Clustering: Supplementary Materials} 
\bigskip

{\large Minjie Wang, Tianyi Yao and Genevera I. Allen}
\end{center}

The supplementary materials are organized as follows. In Appendix~\ref{appA}, we discuss   the algorithm to solve supervised convex biclustering problem. In Appendix~\ref{appB}, we discuss the detail of the exact parameters in the simulation study.

\section{Supervised Convex Biclustering Algorithm}\label{appA}
In this appendix, we discuss the algorithm to solve supervised convex biclustering in Section~\ref{scc-biclust}. The supervised convex biclustering is formulated as:

{
\centering
\scalebox{0.93}{\parbox{1\linewidth}{%
\begin{align*}
\begin{split}
    \minimize_{\mathbf{U}\in\mathbb{R}^{n\times p}, \vect{\theta} \in\mathbb{R}^{n}, \vect{\beta}\in\mathbb{R}^{d}} & \pi_{\mathbf X} \cdot \frac{1}{2} ||\mathbf{X}-\mathbf{U}||_F^2 +  \pi_{\mathbf Y}  \cdot  \ell(\mathbf{y};\vect{\theta}+\mathbf{Z}\vect{\beta}) \\
    & + \lambda \sum_{((i,i'),w_{ii'})\in \mathcal{E}} w_{ii'}\Big|\Big|\left[
\begin{array}{c}
\theta_i  \\
\mathbf{U}_{i\cdot}
\end{array}
\right] - \left[
\begin{array}{c}
\theta_{i'}  \\
\mathbf{U}_{i'\cdot}
\end{array}
\right]\Big|\Big|_2 
 + \lambda \sum_{((j,j'),w_{jj'})\in \mathcal{\tilde E}} \tilde w_{jj'}\Big|\Big| \mathbf{U}_{\cdot j} - \mathbf{U}_{\cdot j'}
 \Big|\Big|_2 .
\end{split}
\end{align*}
}}
\par
}

Notice we cannot use Dykstra-Like Proximal Algorithm (DLPA) mentioned in \citet{weylandt2019dynamic} here as DLPA requires $\ell_2$-type loss whereas our loss $\ell$ is arbitrary here. To address this, we use multi-block ADMM to solve the above problem. To account for the difference between the columns of two centroids, we introduce a new variable $\M$ which is equal to $\U^T$. In this way, the row-wise and column-wise penalty on the difference between two centroids decompose.


We can recast the problem as the equivalent constrained optimization problem:

{
\centering
\scalebox{0.85}{\parbox{1\linewidth}{%
\begin{align*}
    &\minimize_{\mathbf{U}, \vect{\theta}, \vect{\beta},\V,\M} \hspace{2mm} \pi_{\mathbf X} \cdot \frac{1}{2} ||\mathbf{X}-\mathbf{U}||_F^2 +  \pi_{\mathbf Y}  \cdot  \ell(\mathbf{y};\vect{\theta}+\mathbf{Z}\vect{\beta}) + \lambda \underbrace{ \bigg(\sum_{(l,w_{l})\in \mathcal{E}} w_{l}\Big|\Big|{\V}_{\text{row},l.}       \Big|\Big|_2 \bigg)}_{P(\V_{\text{row}};\bw)} + \lambda \bigg( \sum_{(l',\tilde w_{l'})\in \mathcal{\tilde E}} \tilde w_{l'}\Big|\Big|{\V}_{\text{col},l'.}       \Big|\Big|_2\bigg) \\
    & \text{subject to} \hspace{5mm} \D_{\text{row}} \U = \V^{\U}_{\text{row}} , \hspace{5mm} \D_{\text{row}} \vect{\theta} = \V^{\vect{\theta}}_{\text{row}}, \hspace{5mm} \D_{\text{col}}^T \vect{\M} = \V_{\text{col}}, \hspace{5mm} \U^T = \M.
\end{align*}
}}
\par
}

Notice we can rewrite the first two constraints as  $\D_{\text{row}} \begin{bmatrix} \vect{\theta} & \U  \end{bmatrix} - \V_{\text{row}} = \mathbf 0$ where  $\mathbf{V}_{\text{row}} = \begin{bmatrix} \V^{\vect{\theta}}_{\text{row}}  & \V^{\U}_{\text{row}} \end{bmatrix}$. In this way, the augmented Lagrangian is:
\begin{align*}
& \pi_{\mathbf X} \cdot \frac{1}{2} ||\mathbf{X}-\mathbf{U}||_F^2 + \pi_{\mathbf Y}  \cdot  \ell(\mathbf{y};\vect{\theta}+\mathbf{Z}\vect{\beta}) + \lambda \sum_{(l,w_{l})\in \mathcal{E}} w_{l}\Big|\Big|{\V}_{\text{row},l.}       \Big|\Big|_2 + \lambda \sum_{(l',\tilde w_{l'})\in \mathcal{\tilde E}} \tilde w_{l'}\Big|\Big|{\V}_{\text{col},l'.}       \Big|\Big|_2 \\
& + \frac{\rho}{2} \| \D_{\text{row}} \U - \V^{\U}_{\text{row}}  + \Q^{\U}_{\text{row}} \|_F^2 +  \frac{\rho}{2}  \| \D_{\text{row}} \vect{\theta} - \V^{\vect{\theta}}_{\text{row}} + \Q^{\vect{\theta}}_{\text{row}} \|_2^2 + \frac{\rho}{2} \| \D^T_{\text{col}} \M - \V_{\text{col}}  + \Q_{\text{col}} \|_F^2 
\\ &+ \frac{\rho}{2} \| \U^T - \M + \N \|_F^2.
\end{align*}

Here, $\Q$ is the dual variable for $\V$ while $\N$ is the dual variable for $\M$. For each primal and dual variable, the multi-block ADMM has the following updates:
\begin{align*}
\begin{cases}
\U^{(k+1)} &= \hspace{2mm} \argmin \limits_{\U} \pi_{\mathbf X} \cdot \frac{1}{2} ||\mathbf{X}-\mathbf{U}||_F^2 +  \frac{\rho}{2} \| \D_{\text{row}} \U - {\V^{\U}_{\text{row}}}^{(k)}  + {\Q^{\U}_{\text{row}}}^{(k)} \|_F^2  + \frac{\rho}{2} \| \U - {\M^{(k)}}^T + {\N^{(k)}}^T \|_F^2     \\
\vect{\theta}^{(k+1)} &=  \hspace{2mm}  \argmin \limits_{\vect{\theta}}\pi_{\mathbf Y}  \cdot  \ell(\mathbf{y};\vect{\theta}+\mathbf{Z}\vect{\beta}^{(k)})   + \frac{\rho}{2}  \| \D_{\text{row}} \vect{\theta} - {\V^{\vect{\theta}}_{\text{row}}}^{(k)} + {\Q^{\vect{\theta}}_{\text{row}}}^{(k)} \|_2^2 \\
\vect{\beta}^{(k+1)} & =  \argmin \limits_{\vect{\beta}}\pi_{\mathbf Y}  \cdot  \ell(\mathbf{y};\vect{\theta}^{(k+1)}+\mathbf{Z}\vect{\beta})   \\
\M^{(k+1)} &=  \hspace{2mm}  \argmin \limits_{\M} \frac{\rho}{2} \| \D^T_{\text{col}} \M - \V_{\text{col}}^{(k)}  + \Q_{\text{col}}^{(k)} \|_F^2 
+ \frac{\rho}{2} \| {\U^{(k+1)}}^T - \M + \N^{(k)} \|_F^2  \\
\V_{\text{row}}^{(k+1)} & = \argmin \limits_{\V_{\text{row}}} \frac{\rho}{2} \| \D_{\text{row}} \begin{bmatrix} \vect{\theta}^{(k+1)} & \U^{(k+1)}  \end{bmatrix}   - \V_{\text{row}}  + \Q_{\text{row}}^{(k)} \|_F^2  + \lambda \sum_{(l,w_{l})\in \mathcal{E}} w_{l}\Big|\Big|{\V}_{\text{row},l.}       \Big|\Big|_2 \\
\V_{\text{col}}^{(k+1)}   & = \argmin \limits_{\V_{\text{col}}} \frac{\rho}{2} \| \D^T_{\text{col}} \M^{(k+1)} - \V_{\text{col}}  + \Q_{\text{col}}^{(k)} \|_F^2  + \lambda \sum_{(l',\tilde w_{l'})\in \mathcal{\tilde E}} \tilde w_{l'}\Big|\Big|{\V}_{\text{col},l'.}       \Big|\Big|_2 \\
\mathbf{Q}_{\text{row}}^{(k+1)} &= \mathbf{Q}_{\text{row}}^{(k)} + (\mathbf{D}_{\text{row}} \begin{bmatrix} \vect{\theta}^{(k+1)}  &  \mathbf{U}^{(k+1)}   \end{bmatrix}  - \mathbf{V}_{\text{row}}^{(k+1)} )  \\
\mathbf{Q}_{\text{col}}^{(k+1)} &= \mathbf{Q}_{\text{col}}^{(k)} + (\mathbf{D}_{\text{col}}^T \M^{(k+1)}  - \mathbf{V}_{\text{col}}^{(k+1)} ) \\
\N^{(k+1)}  &= \N^{(k)}  + {\U^{(k+1) }}^T - \M^{(k+1)} 
\end{cases}
\end{align*}

Hence, we give Algorithm~\ref{alg:bicscc-alg} to solve  supervised convex biclustering with differentiable loss.  For non-differentiable distance-based loss $\ell$, we can introduce a new block for the non-smooth function $\ell$ and apply multi-block ADMM with simple closed-form solutions for each primal variable update.

\begin{algorithm}[H]
	\caption{Multi-block ADMM algorithm for supervised convex biclustering with differentiable loss $\ell$}
	\label{alg:bicscc-alg}
	\begin{algorithmic}
	
		\WHILE{not converged}

		\STATE $\mathbf{U}^{(k+1)}= (\rho\mathbf{D}_{\text{row}}^T\mathbf{D}_{\text{row}} + \pi_{\mathbf{X}} \cdot \mathbf{I}+ \rho   \mathbf{I} )^{-1}\Big(\pi_{\mathbf{X}}\mathbf{X}+\rho\mathbf{D}_{\text{row}}^T( {\mathbf{V}_{\text{row}}^{\mathbf{U}}}^{(k)} - {\mathbf{Q}_{\text{row}}^{\mathbf{U}}}^{(k)}  )    + \rho ( {\M^{(k)}}^T - {\N^{(k)}}^T)                 \Big)$
		
		\STATE $\vect{\theta}^{(k+1)} = \vect{\theta}^{(k)} - t_k \big(  \pi_{\mathbf{y}} \cdot \nabla \ell(\mathbf{y};\vect{\theta}^{(k)}+\mathbf{Z}\vect{\beta}^{(k)}) + \rho \mathbf{D}_{\text{row}}^T (\mathbf{D}_{\text{row}}\vect{\theta}^{(k)} - {\mathbf{V}_{\text{row}}^{\vect{\theta}}}^{(k)} + {\mathbf{Q}_{\text{row}}^{\vect{\theta}}}^{(k)} ) \big)$
		
		\STATE $\vect{\beta}^{(k+1)}= \vect{\beta}^{(k)} - t_k  \nabla \ell(\mathbf{y}; \vect{\theta}^{(k+1)}+\mathbf{Z}\vect{\beta}^{(k)})$
		
		\STATE $\M^{(k+1)} =  (\mathbf{D}_{\text{col}}\mathbf{D}_{\text{col}}^T +  \mathbf{I}  )^{-1} \bigg(  \mathbf{D}_{\text{col}} ( \V^{(k)}_{\text{col}} - \Q^{(k)}_{\text{col}} ) + {\U^{(k+1)}}^T + \N^{(k)} \bigg)$
		
		\STATE $\mathbf{V}^{(k+1)}_{\text{row}}=\text{prox}_{\lambda/\rho P(\cdot; \bw )} (\mathbf{D}_{\text{row}} \begin{bmatrix} \vect{\theta}^{(k+1)} &  \U^{(k+1)} \end{bmatrix} + \Q_{\text{row}}^{(k)})$
		
		\STATE $\mathbf{V}^{(k+1)}_{\text{col}} =\text{prox}_{\lambda/\rho P(\cdot; \tilde \bw )} (\mathbf{D}_{\text{col}}^T \M^{(k+1)} + \Q_{\text{col}}^{(k)})$
		
		\STATE $\mathbf{Q}_{\text{row}}^{(k+1)} = \mathbf{Q}_{\text{row}}^{(k)} + (\mathbf{D}_{\text{row}} \begin{bmatrix} \vect{\theta}^{(k+1)}  & \mathbf{U}^{(k+1)}  \end{bmatrix}  - \mathbf{V}_{\text{row}}^{(k+1)})$
		
		\STATE $\mathbf{Q}_{\text{col}}^{(k+1)} = \mathbf{Q}_{\text{col}}^{(k)} + (\mathbf{D}_{\text{col}}^T \M^{(k+1)} - \mathbf{V}_{\text{col}}^{(k+1)})$
		
		\STATE  $\N^{(k+1)} = \N^{(k)} + {\U^{(k+1)}}^T - \M^{(k+1)} $

		\ENDWHILE

	\end{algorithmic}
\end{algorithm}

\section{Simulation Setup}\label{appB}

In this appendix, we discuss the detail of the exact parameters for simulating the unlabeled data and supervising auxiliary variables in Section~\ref{sim}.

In the base simulation setup, we consider the case when the supervising auxiliary variable $\y$ is generated from the cluster centroid directly without additional covariates. For each simulation, the data set consists of $n=120$ observations and $p=30$ features with 3 clusters. Each cluster has an equal number of observations for the base simulation. The data is generated from the following model: $\X_{i.} \sim N(\mathbf \mu_k,\sigma^2 \textbf I_p)$, where $i \in G_k$, $k = 1,2,3$ ($G_k$ refers to the observation indices belonging to group $k$).  The supervising auxiliary variable, $y_i$, is generated from different distributions  with parameter $\mathbf \mu_k$ based on data type; the two sources have the shared group label which means $y_i \sim \phi(\mu_k)$, where $i \in G_k$, $k = 1,2,3$ and $\phi$ is a distribution function. We denote $\X_{G_k}$ and $\y_{G_k}$ as the data points and their corresponding supervising auxiliary variable that belong to group $k$. 

We consider two designs of the unlabeled data $\X$: spherical (S) and half-moon (H). In terms of the half moon data, we consider the standard simulated data of three interlocking half moons as suggested by \cite{chi2015splitting} and \cite{wang2019integrative}. For each design, we consider two scenarios where none of the data sources lead to perfect clustering results.

\begin{itemize}
    \item S1: Spherical data: $\X_{G_1}$ and $\X_{G_3}$ overlap, $\X_{G_2}$ are separate from $\X_{G_1}$ and $\X_{G_3}$; $\y_{G_1}$ and $\y_{G_3}$ have two separate clusters, $\y_{G_2}$ are noisy and overlap with $\y_{G_1}$ and $\y_{G_3}$. 
    
    Specifically, $\X_{i.} \sim N(\mathbf \mu_k,\sigma^2 \textbf I_p)$, $i \in G_k$ where $\mathbf  \mu_1 = (1.6\cdot \mathbf 1_{15}^T, 2 \cdot \mathbf 1_{15}^T )^T$, $\mathbf  \mu_2 = (2 \cdot \mathbf 1_{15}^T,  \mathbf 0_{15}^T )^T$, $\mathbf  \mu_3 = (2.4\cdot \mathbf 1_{15}^T, 2 \cdot \mathbf 1_{15}^T )^T$, $\sigma^2 = 1$. For Gaussian supervising auxiliary variable, 
    $y_{i} \sim N(2.25,1)$ for $i \in G_1$; $y_{i} \sim N(4,4)$ for $i \in G_2$; $y_{i} \sim N(5.75,1)$ for $i \in G_3$. For binary supervising auxiliary variable, $y_{i} \sim \text{Bernoulli}(\mu_k)$, $i \in G_k$  where $ \mu_1 = 0.85$, $ \mu_2 = 0.5$, $ \mu_3 = 0.15$. For categorical supervising auxiliary variable, $y_{i} \sim \text{Multinomial}(\mathbf \mu_k)$, $i \in G_k$  where $ \mu_1 = [0.75,0.15,0.1]$, $ \mu_2 = [1/3,1/3,1/3]$, $ \mu_3 = [0.1,0.15,0.75]$. For count-valued  supervising auxiliary variable, $y_{i} \sim \text{Poisson}(1)$ for $i \in G_1$;  $y_{i} \sim \text{Poisson}(9)$ for $i \in G_3$; $y_{i}$ is simulated from a Poisson mixture with $\mu = 1,5,9$ for $i \in G_2$. For survival supervising auxiliary variable, the survival time and censoring indicator are generated with the same censored rate but different hazard rates $\mu_k$s.

    \item S2: Spherical data: $\X_{G_1}$, $\X_{G_2}$ and  $\X_{G_3}$ overlap; $\y$ has three separate  clusters with little overlapping. 
    
    Specifically, $\X_{i.} \sim N(\mathbf \mu_k,\sigma^2 \textbf I_p)$, $i \in G_k$ where $\mathbf  \mu_1 = (-1 \cdot \mathbf 1_{15}^T, \mathbf 0_{15}^T )^T$, $\mathbf  \mu_2 = ( \mathbf 0_{15}^T,  2 \cdot \mathbf 1_{15}^T )^T$, $\mathbf  \mu_3 = (1 \cdot \mathbf 1_{15}^T, \mathbf 0_{15}^T )^T$, $\sigma^2 = 4.4$. For Gaussian supervising auxiliary variable, $y_{i} \sim N(\mu_k,\sigma^2)$, $i \in G_k$ where $ \mu_1 = 1$, $ \mu_2 = 4.5$, $ \mu_3 = 8$. For binary supervising auxiliary variable, it is not possible to simulate binary $\y$ with three separate groups; therefore, we do not include this type of variable in this simulation setup. For categorical supervising auxiliary variable, $y_{i} \sim \text{Multinomial}(\mathbf \mu_k)$, $i \in G_k$  where $ \mu_1 = [0.9,0.05,0.05]$, $ \mu_2 = [0.05,0.9,0.05]$, $ \mu_3 = [0.05,0.05,0.9]$. For count-valued supervising auxiliary variable, $y_{i} \sim \text{Poisson}(\mu_k)$, $i \in G_k$ where $\mu_1 = 1$, $\mu_2 = 10$, $\mu_3 = 23$; For survival supervising auxiliary variable, the survival time and censoring indicator are generated with the same censored rate but different hazard rates $\mu_k$s.

    \item H1: Non-spherical data with three half moons: For the following two scenarios (H1 and H2), we consider the standard simulated data of three interlocking half moons as suggested by \cite{chi2015splitting} and \cite{wang2019integrative}. $\X_{G_1}$ and $\X_{G_3}$ overlap, $\X_{G_2}$ are separate from $\X_{G_1}$ and $\X_{G_3}$; $\y_{G_1}$ and $\y_{G_3}$ have separate two clusters, $\y_{G_2}$ are noisy and overlap with $\y_{G_1}$ and $\y_{G_3}$. The supervising auxiliary variables are simulated similarly as in S1.

    \item H2: Non-spherical data with three half moons: $\X_{G_1}$, $\X_{G_2}$ and  $\X_{G_3}$ overlap; $\y$ has separate three clusters with little overlapping. The supervising auxiliary variables are simulated similarly as in S2.
\end{itemize}

For each of the simulations above, we create a challenging scenario where good clustering results cannot be achieved by clustering either $\X$ or $\y$ alone. 

For the above simulations, we assume that the number of cluster centroids for both sources is the same. Yet, in the case of categorical supervising auxiliary variable, usually, the number of categories we observe in that variable is different from the number of true classes. Hence we consider the following additional simulations. In additional simulation 1 (AS1), we assume the number of clusters of $\X$ is greater than number of categories in $\y$; in AS1, we consider both binary and categorical supervising auxiliary variables. In additional simulation 2 (AS2), we assume the number of categories in $\y$ is greater than number of clusters of $\X$; in AS2, we consider categorical supervising auxiliary variable.

\begin{itemize}

    \item AS1: Categorical/Binary simulation: number of clusters of $\X$ is greater than number of classes of $\y$.
    
    Specifically, for categorical simulation, $\X$ has four clusters with $\X_{i.} \sim N(\mathbf \mu_k,\sigma^2 \textbf I_p)$, $i \in G_k$ where $\mathbf  \mu_1 = (-1 \cdot \mathbf 1_{15}^T, \mathbf 0_{15}^T )^T$, $\mathbf  \mu_2 = (\mathbf 0_{15}^T,  -4 \cdot \mathbf 1_{15}^T )^T$, $\mathbf  \mu_3 = (1 \cdot \mathbf 1_{15}^T, \mathbf 0_{15}^T )^T$, $\sigma^2 = 2$. To make $\X$ has four clusters, we randomly set some of the observations in each group to have different cluster centroids $\tilde \mu_k$ so that those observations form a cluster. The categorical supervising auxiliary variable is simulated similarly as in S2.

    \item AS2: Categorical simulation: number of classes of $\y$ is greater than number of clusters of $\X$.

    Specifically, $\X$ is simulated from three clusters. To make it a challenging scenario, we randomly choose some of the points to be fairly noisy and close to other clusters. The categorical supervising auxiliary variable has five categories: $y_{i} \sim \text{Multinomial}(\mathbf \mu_k)$, $i \in G_k$  where $ \mu_1 = [0.5,0,0,0,0.5]$, $ \mu_2 = [0,0,1,0,0]$, $ \mu_3 = [0,0.5,0,0.5,0]$.

\end{itemize}

Table~\ref{scc-base} shows all the results for the base simulation. Overall we see that our supervised convex clustering outperforms existing methods for different types of supervising auxiliary variables by leveraging information from both sources.

Codes can be found at \url{https://github.com/DataSlingers/SupervisedConvexClustering}.

\end{appendix}

\bibliographystyle{abbrvnat}
\bibliography{main.bib}

\end{document}